\documentclass[apj,twocolappendix,numberedappendix]{emulateapj}
\usepackage{times}
\usepackage{graphicx}
\usepackage{amssymb}
\usepackage{amsmath}
\usepackage{pifont}
\usepackage{graphics}
\usepackage{xcolor}
\usepackage{epsfig}
\usepackage[hidelinks]{hyperref}
\usepackage{amssymb}
\usepackage{courier}

\newcommand{\degrees}{\ensuremath{^{\circ}}}
\newcommand{\chisq}{\ensuremath{\chi^{\,2}}}
\newcommand{\bjdtdb}{\ensuremath{\rm {BJD_{TDB}}}}

\usepackage[normalem]{ulem}
\usepackage{color}

\begin{document}

\title{AstroImageJ: Image Processing and Photometric Extraction for Ultra-Precise Astronomical Light Curves (Expanded Edition)}

\author{
Karen A.\ Collins\altaffilmark{1,2,3}, 
John F.\ Kielkopf\altaffilmark{\,3},
Keivan G.\ Stassun\altaffilmark{1,2}, and
Frederic V.\ Hessman\altaffilmark{\,4}
}
\altaffiltext{1}{Department of Physics \& Astronomy, Vanderbilt University, Nashville, TN 37235, USA, karenacollins@outlook.com}
\altaffiltext{2}{Department of Physics, Fisk University, Nashville, TN 37208, USA}
\altaffiltext{3}{Department of Physics \& Astronomy, University of Louisville, Louisville, KY 40292, USA}
\altaffiltext{4}{Inst. f. Astrophysik, Georg-August-Universit\"{a}t G\"{o}ttingen}

\begin{abstract}
ImageJ is a graphical user interface (GUI) driven, public domain, Java-based, software package for general image processing traditionally used mainly in life sciences fields. The image processing capabilities of ImageJ are useful and extendable to other scientific fields. Here we present AstroImageJ (AIJ), which provides an astronomy specific image display environment and tools for astronomy specific image calibration and data reduction. Although AIJ maintains the general purpose image processing capabilities of ImageJ, AIJ is streamlined for time-series differential photometry, light curve detrending and fitting, and light curve plotting, especially for applications requiring ultra-precise light curves (e.g., exoplanet transits). AIJ reads and writes standard FITS files, as well as other common image formats, provides FITS header viewing and editing, and is World Coordinate System (WCS) aware, including an automated interface to the astrometry.net web portal for plate solving images. AIJ provides research grade image calibration and analysis tools with a GUI driven approach, and easily installed cross-platform compatibility.  It enables new users, even at the level of undergraduate student, high school student, or amateur astronomer, to quickly start processing, modeling, and plotting astronomical image data with one tightly integrated software package. 
\end{abstract}


\section{Introduction}
\label{sec:intro}

In many areas of astronomy, there is a need for image processing and analysis capabilities and light curve extraction. One such general purpose environment is IRAF \citep{Tody:1986,Tody:1993}\footnote{IRAF is distributed by the National Optical Astronomy Observatories, which are operated by the Association of Universities for Research in Astronomy, Inc., under cooperative agreement with the National Science Foundation.}. However, especially for ultra-precise photometry in fast-paced areas of research such as exoplanet transits and microlensing, there remains a need for a general, off-the-shelf integrated analysis environment that is at once sophisticated yet easy to use. Indeed, citizen science and professional-amateur collaborations increasingly require robust tools that can deliver research-grade results while enabling broad usability.

Here we present AstroImageJ (AIJ), an astronomical image analysis software package that is based on ImageJ (IJ; \citealt{Rasband:1997}), but includes customizations to the base IJ code and a packaged set of software plugins that provide an astronomy specific image display environment and tools for astronomy specific data reduction, analysis, modeling, and plotting. AIJ and IJ are public domain, open source, Java programs inspired by NIH Image for the Macintosh computer. Some AIJ capabilities were derived from the \textit{Astronomy} plugins package\footnote{\url{http://www.astro.physik.uni-goettingen.de/~hessman/ImageJ/Astronomy/}}. Some astronomical algorithms are based on code from \textit{JSkyCalc} written by John Thorstensen of Dartmouth College. Because AIJ is Java code, the package is compatible with computers running Apple OS X, Microsoft Windows, and the Linux operating system (OS).

AIJ is a general purpose astronomical image processing tool, plus it provides interfaces to streamline the interactive processing of image sequences. The current release (version 3.2.0) includes the following features and capabilities, where (I) indicates a feature provided by the underlying ImageJ platform, (I+) indicates an ImageJ feature that has been improved, (A+) indicates a feature based on the \textit{Astronomy} plugins package, but with significant new capabilities, and (N) indicates a new feature that is available in AIJ, but not available in ImageJ or the \textit{Astronomy} plugins package:

\begin{itemize}
  \item (N) Interactive astronomical image display supporting multiple image stacks with fast image zooming and panning, high-precision contrast adjustment, and pixel data display similar to \textit{SAOImage DS9} \citep{DS9:2000}\footnote{SAOImage DS9 has been made possible by funding from the Chandra X-ray Science Center (CXC) and the High Energy Astrophysics Science Archive Center (HEASARC).}
  \item (N) Live mouse pointer photometer
  \item (N) Sky orientation of image and pixel scale display in a non-destructive image overlay
  \item (A+) Reads and writes FITS images with standard headers, as well as most other common image formats (e.g. tiff, jpeg, png, etc.)
  \item (N) Data Processor facility for image calibration including bias, dark, flat, and non-linearity correction with an option to run in real-time during observations
  \item (A+) Interactive time-series multi-aperture differential photometry with detrend parameter extraction, and an option to run in real-time during observations
  \item (N) Photometric uncertainty calculations including source and sky Poisson noise, dark current, detector readout noise, and quantization noise, with automatic propagation of single-aperture uncertainties through differential photometry, normalization, and magnitude calculations 
  \item (N) Comparison star ensemble changes without re-running differential photometry 
  \item (N) Interactive multi-curve plotting streamlined for display of light curves 
  \item (N) Interactive light curve fitting with simultaneous detrending 
  \item (N) Plate solving and addition of WCS headers to images seamlessly using the Astrometry.net web interface
  \item (N) Time and coordinate conversion with capability to update/enhance FITS header content (airmass, \bjdtdb, etc.)
  \item (A+) Point and click radial profile (i.e.\ seeing profile) plots
  \item (N) FITS header viewing and editing
  \item (N) Astronomical coordinate display for images with WCS 
  \item (N) Object identification via an embedded SIMBAD interface 
  \item (A+) Image alignment using WCS headers or apertures to correlate stars
  \item (N) Non-destructive object annotations/labels using FITS header keywords
  \item (I) Mathematical operations of one image on another or an image stack, and mathematical and logical operations on single images or image stacks
  \item (I+) Color image creation
  \item (N) Optionally enter reference star apparent magnitudes to calculate target star magnitudes automatically
  \item (N) Optionally create Minor Planet Center (MPC) format for direct submission of data to the MPC 
\end{itemize}

AIJ is currently used by most of the $\sim 30$ member Kilo-degree Extremely Little Telescope (KELT; \citealt{Pepper:2003,Pepper:2007}) transit survey photometric follow-up team, 
so far resulting in 10 planets published
\citep{Siverd:2012,Beatty:2012,Pepper:2013,Collins:2014,Bieryla:2015,Fulton:2015,Eastman:2016,Rodriguez:2016,Kuhn:2016}, 
and at least 8 more in press or in preparation as of this writing. 
AIJ users on the team include amateur astronomers, undergraduate and graduate students, and professional astronomers. AIJ is also used by the KELT science team to optimize the precision of, and determine the best detrending parameters for, all follow-up light curves that are included in the analysis of new planet discoveries. AIJ is deployed in multiple undergraduate university teaching labs and is also used to teach exoplanet transit analysis to high school students. We and the KELT follow-up team have verified the accuracy of AIJ against a number of traditional scientific and commercial photometric extraction packages, including IRAF, IDL\footnote{http://www.harrisgeospatial.com; IDL is a product of Exelis Visual Information Solutions, Inc., a subsidiary of Harris Corporation.}\textsuperscript{,}\footnote{http://idlastro.gsfc.nasa.gov}, and MaxIm DL\footnote{http://www.cyanogen.com}. The IRAF and IDL photometric capabilities were adapted from DAOPHOT \citep{Stetson:1987}. We do not track the number of AIJ downloads, but we estimate that there are several hundred active AIJ users based on AIJ user forum\footnote{\label{userforum}\url{http://astroimagej.1065399.n5.nabble.com/}} statistics.

AIJ's ultra-precise photometric capabilities are demonstrated by \citet{Collins:2017a}, where they achieved an RMS of 183 and 255 parts per million for the transit model residuals of the combined and five minute binned ground-based light curves of WASP-12b and Qatar-1b, respectively, and transit timing residuals from a linear ephemeris of less than $\sim30$~s. These results are enabled by a multi-star photometer that allows fixed or variable radius apertures, a variety of options to calculate sky-background, including sky-background star rejection, and high precision centroid capabilities, including the ability to properly centroid on defocused stars. In addition, AIJ's interactive GUIs and tightly coupled extraction of differential photometry and detrend parameters, light curve plotting, comparison star ensemble manipulation, and simultaneous fitting of the data to a transit model and detrending parameters, enable the user to quickly optimize detrended and fitted light curve precision. 

For example, stars can be added to or removed from the comparison ensemble (without re-running photometry) and detrending parameters can be changed instantly by clicking to enable or disable each one. When a change is made, the light curve and fitted model plots are automatically updated and statistical values indicating the goodness of the model fit, such as RMS and the Bayesian Information Criterion, are instantly updated. These interactive features enable a user to quickly determine the best aperture settings, comparison ensemble, and detrend parameter set. 

Finally, if AIJ is operated in ``real-time'' mode during time-series observations, the images are calibrated, photometry is extracted, and data are plotted, detrended, and model fitted automatically as images are written to the local system's hard disk by any camera control software package. This capability works independent of (and does not interfere with) an observatory's telescope and camera control software and allows the user to explore exposure time, defocus, aperture settings, and comparison star ensemble to ensure high-precision photometric results in the final post-processed data. 

The following sections provide detailed descriptions of the astronomy specific capabilities of AIJ, as well as detailed descriptions of the AIJ user interface. A peer reviewed abridged version of this work is also available \citep{Collins:2017b}. The AIJ User Guide, installation packages, and installation instructions are available for download at the AIJ website\footnote{\url{http://www.astro.louisville.edu/software/astroimagej}}. Most of the AIJ user interface panels include ``tool-tip'' help messages that optionally pop up when the mouse pointer is positioned over an item in the display for more than a second. An AIJ user forum\textsuperscript{\ref{userforum}} is available to facilitate shared support for the software. AIJ inherits all of the basic image manipulation and analysis functionality from IJ. The IJ website\footnote{\url{http://imagej.nih.gov/ij/}} provides detailed user guides and descriptions of its functionality.  

\section{AIJ Overview and Basic Capabilities}

\subsection{Toolbar}

\begin{figure*}
\begin{center}
\resizebox{\textwidth}{!}{
\includegraphics*{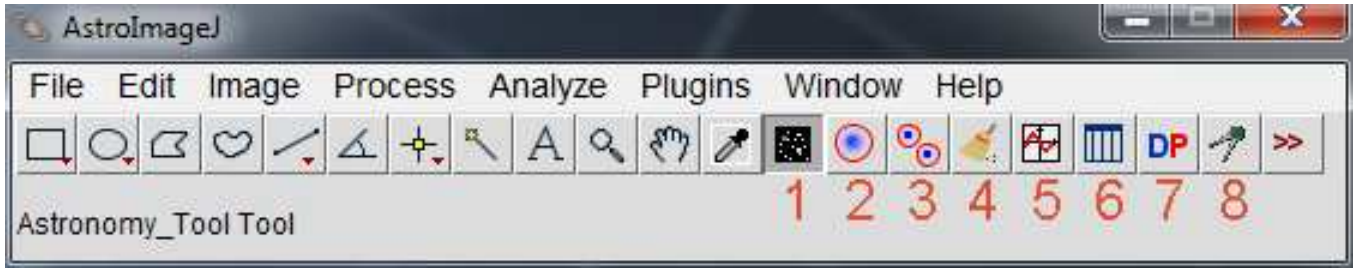} }
\caption[AIJ Toolbar]{The AIJ Toolbar. The icon shown depressed and labeled as 1 indicates that AIJ is in astronomy mode. In this mode, all images open into the Astronomical Image Display mode discussed in \S \ref{sec:aijdisplay}. Icon 2 starts the single aperture photometry mode discussed in \S \ref{sec:singleaperture}. A double-click on Icon 2 opens the \textit{Aperture Photometry Settings} panel discussed in Appendix \ref{sec:photset}. Icon 3 starts the Multi-Aperture photometer module discussed in \S \ref{sec:multiaperture}. Icon 4 clears all labels and apertures from the image display. Icon 5 starts the Multi-Plot module discussed in \S \ref{sec:multiplot}. Icon 6 opens previously saved photometry ``measurements tables'' (see Appendix \ref{app:measurementstable}). Icon 7 opens the \textit{Data Processor} panel discussed in \S \ref{sec:dataprocessor}. Icon 8 opens the \textit{Coordinate Converter} panel discussed in Appendix \ref{sec:coordinateconverter}.}
\label{fig:aijtoolbar}
\end{center}
\end{figure*}

When AIJ is started, the AIJ Toolbar opens and presents the eight AIJ-specific toolbar icons labeled as 1-8 in Figure \ref{fig:aijtoolbar}. Each of those icons provides direct access to an AIJ analysis tool or function. The icon shown depressed and labeled as 1 $\left(\,\includegraphics[scale=0.45, trim=1.0mm 1.8mm 1.3mm 0mm]{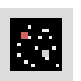}\,\right)$ indicates that AIJ is in astronomy mode. In this mode, all images open into the astronomical image display mode discussed in \S \ref{sec:aijdisplay}. Icon 2 $\left(\,\includegraphics[scale=0.38, trim=1.7mm 2.1mm 1.3mm 0mm]{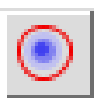}\,\right)$ starts single aperture photometry mode as discussed in \S \ref{sec:singleaperture}. A double-click on Icon 2 opens the \textit{Aperture Photometry Settings} panel discussed in Appendix \ref{sec:photset}. Icon 3 $\left(\,\includegraphics[scale=0.38, trim=1.5mm 2.5mm 1.7mm 0mm]{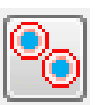}\,\right)$ starts the Multi-Aperture differential photometry module discussed in \S \ref{sec:multiaperture}. Icon 4 $\left(\,\includegraphics[scale=0.33, trim=1.4mm 2.1mm 1.4mm 0mm]{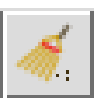}\,\right)$ clears all labels and apertures from the image display. Icon 5 $\,\left(\includegraphics[scale=0.47, trim=0mm 1.5mm 1.2mm 0mm]{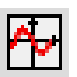}\,\right)$ starts the Multi-Plot module discussed in \S \ref{sec:multiplot}. Icon 6 $\left(\,\includegraphics[scale=0.37, trim=1.0mm 2.1mm 1.3mm 0mm]{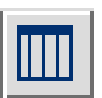}\,\right)$ opens previously saved photometry ``measurements tables'' (see Appendix  \ref{app:measurementstable}). Icon 7 $\left(\,\includegraphics[scale=0.37, trim=1.2mm 2mm 1.4mm 0mm]{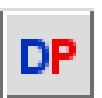}\,\right)$ opens the \textit{Data Processor} panel discussed in \S \ref{sec:dataprocessor}. Icon 8 $\left(\,\includegraphics[scale=0.37, trim=1.2mm 2.2mm 1.4mm 0mm]{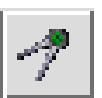}\,\right)$ opens the \textit{Coordinate Converter} panel discussed in Appendix \ref{sec:coordinateconverter}. 

The 12 icons to the left of the AIJ icons and all of the menu options are standard IJ tools. These tools can also be used in AIJ, but normally only the \textit{File} menu options are used for typical time-series data reductions. A single image is opened from the AIJ Toolbar using \textit{File}$\rightarrow$\textit{Open}. A time-series of images is opened into an image ``stack'' from the AIJ Toolbar using \textit{File}$\rightarrow$\textit{Import}$\rightarrow$\textit{Image Sequence}. Alternatively, an image or an image sequence can be opened by dropping the file or folder, respectively, onto the bottom area of the AIJ Toolbar, or OS options can be enabled to automatically open images into AIJ in response to a double click on the file in an OS window. If all images in a sequence will not fit into the computer memory allocated to AIJ, the sequence can be opened as a ``virtual stack''. In this mode, the stack of images can be processed as if all images exist in memory, but AIJ loads only the single active/displayed image into memory. Virtual stacks perform more slowly than standard stacks, but memory requirements are minimal. All AIJ settings are persistent across sessions. Settings for specific configurations can be saved and reloaded later as needed. 

\subsection{Astronomical Image Display}\label{sec:aijdisplay}

\begin{figure*}
\begin{center}
\resizebox{0.85\textwidth}{!}{
\includegraphics*{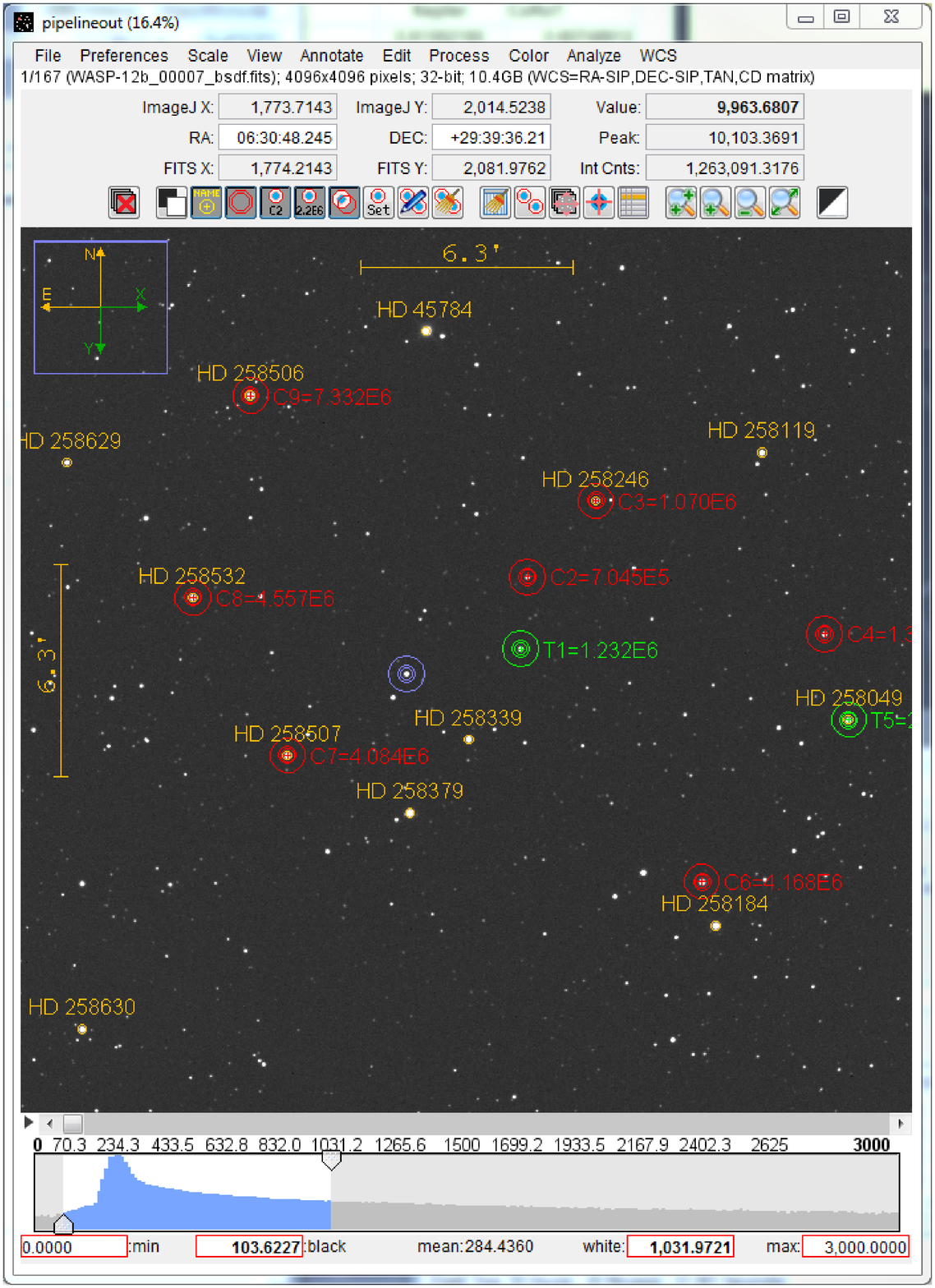} }
\caption{The AIJ image display. A wide range of astronomy specific image display options and image analysis tools are available from the menus, quick access icons, and interactive histogram. See text for details.}
\label{fig:aijimagedisplay}
\end{center}
\end{figure*}

Many popular image file formats are supported by AIJ, including the Flexible Image Transport System (FITS; \citealt{Wells:1981,Pence:2010}) file format. The astronomical image display shown in Figure \ref{fig:aijimagedisplay} is unique to AIJ and offers numerous display options useful to astronomers. An image can also be displayed in plain IJ display mode, which has no contrast controls, analysis controls, or mouse pointer data decorating the image, by disabling the option at \textit{Preferences}$\rightarrow$\textit{Use astro-window when images are opened} in the menus above an open image. The setting will apply to any new images that are opened. To return to astronomical image display mode, click on the \textit{astronomy mode} icon $\left(\,\includegraphics[scale=0.45, trim=1.0mm 1.8mm 1.3mm 0mm]{aijastromodeicon}\,\right)$ in the AIJ toolbar. Then left-click in an open image. To make the change permanent, re-enable the option at \textit{Preferences}$\rightarrow$\textit{Use astro-window when images are opened}.

In astronomy mode, a menu system is available at the top of the image display window to provide access to all astronomy specific AIJ features. A row of quick access icons for control of frequently used image display options and image analysis tools is located directly above the image. Pixel and World Coordinate System (WCS; \citealt{Greisen:2002,Calabretta:2002,Greisen:2006}) information describing the image location pointed to by the mouse cursor is displayed in the three rows above the quick access icons. Image and WCS format information is displayed in the space under the image menus. A non-destructive image overlay optionally displays active apertures (green = target, red = comparison), object annotations, plate scale, and image orientation on the sky. 

The \textit{Scale} menu above an image display offers options for the control of image brightness and contrast (i.e. image scale). By default, image scale is set automatically and linearly maps the pixel values in the range $mean-0.5\sigma$ through $mean+2\sigma$ to 256 shades of gray running from black through white. The negative image icon $\left(\,\includegraphics[scale=0.37, trim=1.5mm 2.0mm 1.7mm 0mm]{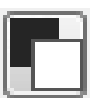}\,\right)$ reverses the mapping to white through black. Several alternate mappings are available at the AIJ Toolbar menu item \textit{Image}$\rightarrow$\textit{Lookup Tables}. The auto scale range can be customized at \textit{Scale}$\rightarrow$\textit{set auto brightness \& contrast parameters}. Image scale can be manually controlled using the interactive histogram and direct entry fields under the image. The histogram region highlighted in blue indicates the range of pixel values currently being mapped to black through white (by default) in the image display. The controls on the left and right of the blue highlighted region can be dragged left and right, or the entire blue region can be dragged, to change the image scale. The image updates in real time. The four boxes under the histogram labeled \textit{min}, \textit{black}, \textit{white}, and \textit{max} indicate the minimum pixel value displayed in the histogram, the pixel value mapped to black in the image, the pixel value mapped to white in the image, and the maximum pixel value displayed in the histogram. By default, the \textit{min} and \textit{max} values are set to the minimum and maximum pixel values in the displayed image. To provide finer control of the image scale settings, the \textit{min} and \textit{max} histogram values can be manually set by the user after enabling \textit{Scale}$\rightarrow$\textit{Use fixed min and max histogram values}. A red border indicates which boxes are available for user input. The \textit{$<$Enter$>$} key \textit{must} be pressed to activate a new user entered value. The auto scale icon above an image $\left(\,\includegraphics[scale=0.37, trim=1.5mm 2.6mm 1.7mm 0mm]{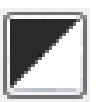}\,\right)$ restores the auto scale settings. If \textit{Scale}$\rightarrow$\textit{auto brightness and contrast} is enabled, a change to a displayed image or a change of which image is displayed from a stack resets the image scale to the automatic settings. To retain manually set values of \textit{black} and \textit{white}, enable \textit{Scale}$\rightarrow$\textit{fixed brightness and contrast}. To automatically set \textit{min} and \textit{black} to the minimum pixel value and \textit{white} and \textit{max} to the maximum pixel value, select \textit{Scale}$\rightarrow$\textit{full dynamic range}. Image scale may also be adjusted by a right-click and drag in the image. Dragging up and down changes the brightness. Dragging left and right changes the contrast. 

The zoom setting of an image display is most easily changed by rolling a mouse wheel, if available, to zoom the image in and out at the mouse pointer. Alternatively, image zoom changes, centered at the last clicked location in an image, can be controlled using the four magnifying glass icons on the right side of the quick access row, or by pressing the \textit{up} and \textit{down} arrows on the keyboard. Zoom can also be changed with a control-left-click or control-right-click in the image to zoom in or out, respectively, at the clicked point. Image pan is controlled by a left-click and drag in the image. By default, a middle-click in an image centers the clicked location in the image display. However, this feature can be disabled using the menu option \textit{Preferences}$\rightarrow$\textit{Middle click centers image at clicked position}. Right-click and drag in an image reports arc-length in place of integrated counts in the lower-right-hand box above an image. Alt-left-click near an object to produce an azimuthally averaged radial profile (i.e. a seeing profile) plot. See Appendix \ref{sec:radialprofile} for more radial profile plot details.

The \textit{View} menu above an image provides settings to invert the display of an image in \textit{x} and/or \textit{y}. The \textit{View} menu also provides options to enable or disable the display of the zoom indicator, the \textit{X}, \textit{Y}, \textit{N}, and \textit{E} directional arrows, and the image plate scale indicators in the image overlay. If WCS header information is available, AIJ automatically calculates the $x$- and $y$-axis plate scales and the orientation of the image on the sky. If no WCS information is available, the image scale and orientation on the sky can be set manually at \textit{WCS}$\rightarrow$\textit{Set pixel scale for images without WCS} and \textit{WCS}$\rightarrow$\textit{Set north and east arrow orientations for images without WCS}, respectively. Image scale and orientation can be extracted from an image with WCS headers and stored in the manual settings by using the menu option \textit{WCS}$\rightarrow$\textit{Save current pixel scale and image rotation to preferences}. This option is useful for cases where at least one image in a time-series has been plate-solved, but others have not.

The blue aperture shown near the center of Figure \ref{fig:aijimagedisplay} moves with the mouse pointer. As the mouse pointer is moved around in the image, the value of the pixel at the mouse pointer and the peak pixel value and background-subtracted integrated counts (see \S \ref{sec:singleaperture}) within the mouse pointer aperture are updated in the right-hand column of data displayed above the image. This interactive mouse photometer helps to quickly assess which stars are suitable comparison stars during differential photometry set up. When AIJ is used in real-time data reduction mode (see \S \ref{sec:dataprocessor}), the mouse photometer helps to quickly determine an appropriate image exposure time and telescope defocus setting.

If a time-series of images is opened as an image stack, a scroll bar is displayed directly under the image as shown in Figure \ref{fig:aijimagedisplay}. The scroll bar can be moved left and right, or the \textit{left} and \textit{right} arrow keys on the keyboard can be used to display different images from the sequence. The right-pointing ``play'' icon to the left of the scroll bar will animate the image sequence at a predefined rate. The rate is set by right-clicking on the play icon.  

The annotation feature allows objects to be labeled non-destructively in an image overlay. New object annotations can be added manually, or if the image has WCS information, target names can be extracted from SIMBAD and displayed by right-clicking on an object. Annotations can be edited or deleted by right-clicking near the annotated object (i.e. below and near the center of the annotation text). The annotations can also be stored in the FITS header for display at a later time. The \textit{Annotate} menu provides annotation and related FITS header storage options. The annotate icon above an image $\left(\,\includegraphics[scale=0.37, trim=1.5mm 1.9mm 1.7mm 0mm]{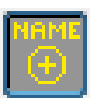}\,\right)$ toggles the display of annotations in the image overlay.

The second through fifth icons shown as depressed in Figure \ref{fig:aijimagedisplay} control which components of apertures (see \S \ref{sec:singleaperture}) are displayed and whether an aperture is to be centroided on the nearest star when it is placed. The delete slice icon $\left(\,\includegraphics[scale=0.38, trim=2mm 2.1mm 2mm 0mm]{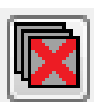}\,\right)$ removes the currently displayed slice from a stack, but does not remove the image from permanent storage. The clear overlay icon $\left(\,\includegraphics[scale=0.37, trim=1.5mm 2.35mm 1.7mm 0mm]{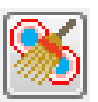}\,\right)$ clears all apertures and annotations from the image overlay.  The show apertures icon $\left(\,\includegraphics[scale=0.37, trim=1.5mm 2mm 1.5mm 0mm]{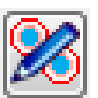}\,\right)$ displays stored apertures (see \S \ref{sec:multiaperture}) in the image overlay. The clear measurements icon $\left(\,\includegraphics[scale=0.37, trim=1.5mm 2.3mm 1.5mm 0mm]{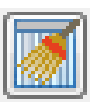}\,\right)$ clears all data from the measurements table (see Appendix \ref{app:measurementstable}). The Multi-Aperture $\left(\,\includegraphics[scale=0.37, trim=1.5mm 2.4mm 1.5mm 0mm]{aijmultiapertureicon}\,\right)$, Stack-Aligner $\left(\,\includegraphics[scale=0.37, trim=1.5mm 2.1mm 1.5mm 0mm]{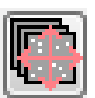}\,\right)$, Astrometry $\left(\,\includegraphics[scale=0.37, trim=1.5mm 2.1mm 1.3mm 0mm]{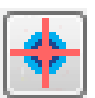}\,\right)$, and FITS header editor $\left(\,\includegraphics[scale=0.37, trim=1.5mm 2.1mm 1.5mm 0mm]{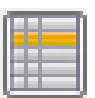}\,\right)$ icons provide access to features described in \S \ref{sec:multiaperture}, Appendix \ref{sec:stackaligner}, Appendix \ref{sec:astrometry}, and Appendix \ref{sec:fitsheadereditor}, respectively.

\subsection{Utilities, Algorithms, and Measurements Tables}
In Appendix \ref{app:utilities} we describe the following integrated utilities: Coordinate Converter (\ref{sec:coordinateconverter}), FITS Header Editor (\ref{sec:fitsheadereditor}), Astrometry/Plate Solving (\ref{sec:astrometry}), Image Alignment (\ref{sec:stackaligner}), Radial Profile plotting (\ref{sec:radialprofile}), Photometry Settings (\ref{sec:photset}), Data Processor FITS Header Updates (\ref{sec:fitsheaderupdates}), and Save All (\ref{sec:saveall}).

In Appendix \ref{app:photerr} we describe photometric uncertainty calculations. In Appendix \ref{app:apparentmagnitude} we describe the optional apparent magnitude and apparent magnitude uncertainty calculations. In Appendix \ref{app:measurementstable} we describe the measurements table used to store photometric results. 

\section{Data Processor: Image Calibration and Reduction}\label{sec:dataprocessor}

\begin{figure*}
\begin{center}
\resizebox{\textwidth}{!}{
\includegraphics*{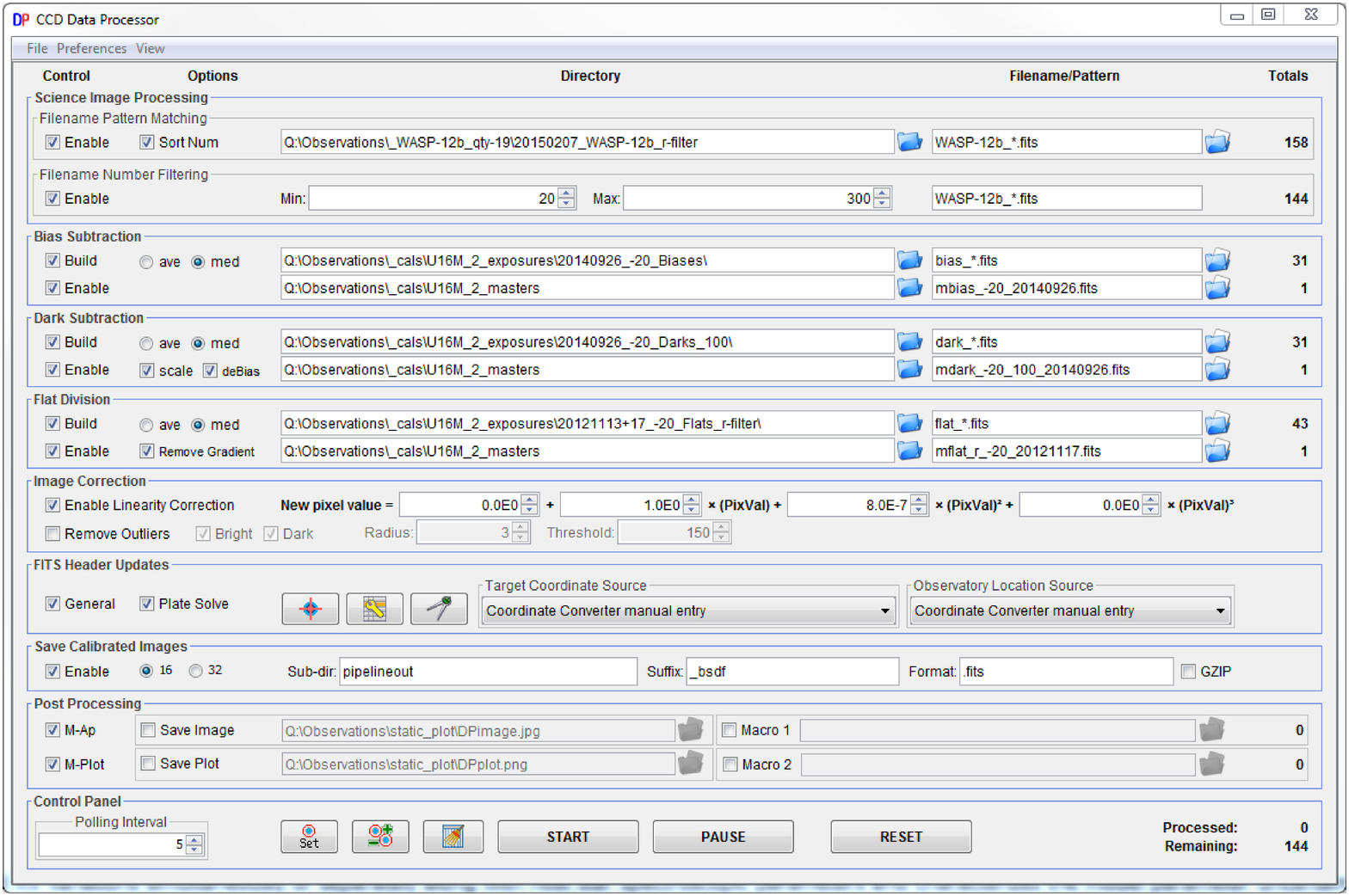} }
\caption{The \textit{Data Processor} panel. \textit{Data Processor} automates the building of master calibration files, calibration of a time series of images, and optionally performs differential photometry and light curve plotting. See text for more details.}
\label{fig:aijdataprocessor}
\end{center}
\end{figure*}

The Data Processor (DP) module provides tools to automate the build of master calibration images, automate calibration of image sequences, and optionally perform differential photometry and light curve plotting. The DP module is started by clicking the DP icon  $\left(\,\includegraphics[scale=0.37, trim=1.3mm 2.1mm 1.3mm 0mm]{aijdpicon}\,\right)$ on the AIJ Toolbar (labeled 7 in Figure \ref{fig:aijtoolbar}). The user interface is shown in Figure \ref{fig:aijdataprocessor}. DP operates much like a script in that it processes selected calibration and science images in a user defined manner. Fields are provided to define the directory/folder locations and filename patterns of data to be processed. Checkboxes are provided to enable various tasks that can be included in the data processing session. Disabling certain checkboxes will automatically disable other related input fields as appropriate to help the user understand which input fields are interconnected. 

The path and image filename pattern matching the raw science images to be processed are set in the \textit{Science Image Processing} sub-panel. File paths and names can be dragged and dropped from the OS into a field of the DP panel to minimize typing. The number of files matching the pattern at the specified path shows in the right-hand column labeled \textit{Totals}. The science images can be further filtered based on the image sequence numbers in the filename by entering minimum and/or maximum numbers in the second row of the sub-panel.

Master bias, dark, and flat images can be created by enabling the \textit{Build} options and entering the paths and filename patterns matching the raw calibration images on the \textit{Build} lines of the \textit{Bias Subtraction}, \textit{Dark Subtraction}, and \textit{Flat Division} sub-panels, respectively. The master calibration files are saved using the path and filename specifications in the \textit{Enable} lines of the three sub-panels. As with the science images, the number of files matching the pattern for each raw calibration image set shows in the right-hand column labeled \textit{Totals}. If one or more of the master calibration files have been previously built, the \textit{Build} option in the respective sub-panel(s) can be disabled to skip those master build steps. In this case, the master calibration path and filename should be specified on the \textit{Enable} line of the respective sub-panel. AIJ provides the option to either average or median combine the raw images when producing the master calibration files. If the dark subtraction \textit{scale} option is enabled, the master dark pixel values are scaled by to the ratio of the science image exposure time to the master dark image representative exposure time. The dark subtraction \textit{deBias} option controls whether the master dark is saved with or without the bias image subtracted, and specifies whether the master dark image being used for dark subtraction was saved with or without the bias image subtracted. The dark \textit{scale} and \textit{deBias} options are available only if bias subtraction is enabled.

Bias subtraction, dark subtraction, and flat-field division can be individually selected using the \textit{Enable} option in the corresponding sub-panel. The master calibration image paths and filenames are specified on the \textit{Enable} line of each sub-panel. When a file path has been entered correctly and matches a stored file, a count of ``1'' will appear in the \textit{Totals} column. 

The \textit{Image Correction} sub-panel provides the option to implement CCD nonlinearity correction. This option replaces each pixel's ADU value in the bias-subtracted dark, flat, and science images with the corrected ADU value
\begin{equation}
\rm{ADU_{corrected}}=c_0 + c_1\rm{ADU} + c_2\rm{ADU}^2  + c_3\rm{ADU}^3,
\label{eq:nonlinearity}
\end{equation}
\noindent where the coefficients $\rm{c_n}$ describe the non-linear behavior of the detector. Bias subtraction must be enabled and a master bias image specified to enable the use of non-linearity correction. The \textit{Remove Outliers} option uses thresholded median filtering to remove artifacts from science images. This option is useful for improving the cosmetic appearance of images (e.g. to clean cosmic ray hits and/or hot and cold pixels). However, we strongly suggest that this option \textit{not} be used when calibrating images for photometric extraction since it can unpredictably impact the results. 

The \textit{FITS Header Updates} sub-panel \textit{General} option enables the calculation of new astronomical data (e.g. airmass, time in \bjdtdb, target altitude, etc.) and the addition of those data to the calibrated science image's FITS header  (see Appendix \ref{sec:fitsheaderupdates} for set up details). The \textit{Plate Solve} option enables astrometry and the addition of the resulting WCS data to the calibrated (and optionally to the raw) science image's FITS header. Click the \textit{Astrometry Settings} icon $\left(\,\includegraphics[scale=0.34, trim=1.7mm 2.8mm 1.7mm 0mm]{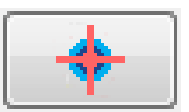}\,\right)$ in the  \textit{FITS Header Updates} sub-panel to access the astrometry set up options discussed in Appendix \ref{sec:astrometry}.

The \textit{Save Calibrated Images} sub-panel provides file format and file naming options for saving calibrated images. The \textit{16} and \textit{32} options specify that the calibrated images be saved in 16-bit integer or 32-bit floating point format, respectively. The \textit{Sub-dir} box allows the specification of an optional subdirectory of the science image directory to be used when saving the calibrated science images. The \textit{Suffix} box allows the specification of an optional suffix to be appended to the raw science image file name when saving calibrated science images. The \textit{Format} box allows the specification of a file format to use when saving the calibrated science images. If this box is left blank, the raw science image format is used. The \textit{Format} box tool-tip shows all available file formats.

The \textit{Post Processing} sub-panel provides options to run Multi-Aperture (see \S \ref{sec:multiaperture}) and Multi-Plot (see \S \ref{sec:multiplot}) after each image is calibrated to perform differential photometry and display a light curve as the image data are processed. This feature is particularly useful for real-time reduction of data at the telescope. Other options allow the current light curve plot and image display to be written to a file after each science image is calibrated. These images can be used to update websites to show the progress of observations.

The \textit{Control Panel} sub-panel provides control over the master calibration build and image calibration process. If the \textit{Polling Interval} is set to zero, a click on the \textit{START} button will build any specified calibration files and process all specified science images. Then DP will return to the stop state. This mode is ideal for post-observation calibration of data. To calibrate data in real-time, the \textit{Polling Interval} is set to a positive number of seconds $t$. After clicking on \textit{START} in this mode, all science images matching the file name and number filters will be processed, and then DP will search for new files matching the pattern every $t$ seconds. If DP is operating in this mode while observations are underway, calibration of a new image written from the camera will be started within $t$ seconds, and optionally processed by Multi-Aperture and Multi-Plot to update the light curve plot. After \textit{START} has been clicked, processing can be suspended by clicking on the \textit{PAUSE} button. In pause mode, the \textit{START} button is labeled \textit{CONTINUE}. A click on the \textit{CONTINUE} button resumes processing, or a click on the \textit{RESET} button returns DP back to the original state. Another click on the \textit{START} button will begin processing \textit{all} specified data again. 

A click on the \textit{Aperture Settings} button $\left(~\includegraphics[scale=0.35, trim=2.3mm 2.5mm 2.5mm 0mm]{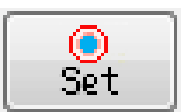}~\right)$ in the \textit{Control Panel} sub-panel opens the \textit{Aperture Photometry Settings} panel discussed in Appendix \ref{sec:photset}. A click on the \textit{Change Apertures} button $\left(~\includegraphics[scale=0.34, trim=2.3mm 2.8mm 2.5mm 0mm]{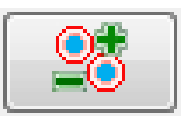}~\right)$ causes the \textit{Multi-Aperture Measurements} set up panel discussed in \S \ref{sec:multiaperture} to open the next time the \textit{START} button is pressed. If the \textit{Change Apertures} button is not clicked after the \textit{RESET} button is clicked, the same apertures and aperture locations from the previous DP run will be used the next time the \textit{START} button is clicked (i.e. the \textit{Multi-Aperture Measurements} set up panel will \textit{not} open before multi-aperture photometry starts). The \textit{Clear Measurements Table Data} button $\left(~\includegraphics[scale=0.34, trim=2.3mm 2.8mm 2.5mm 0mm]{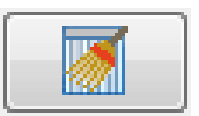}~\right)$ allows the user to clear all data from the measurements table at any time. If the \textit{Clear Measurements Table Data} button is not clicked before starting a new DP run, new measurements will be appended to any existing measurements in the table. The number of images processed and the number of images remaining to be processed are displayed in the \textit{Totals} column of the \textit{Control Panel} sub-panel. 

A detailed log of all processing steps with timestamps is created by default. The log function can be disabled by deselecting the DP menu item \textit{View}$\rightarrow$\textit{Show log of data processing history}. The timestamps can be enabled or disabled in the same menu.

\section{Ultra-Precise Photometry and Light Curve Capabilities}\label{sec:photometry}

AIJ provides interactive interfaces for single aperture photometry and multi-aperture differential photometry. The differential photometry interface is designed to automatically process a time-series of images and measure the light curves of exoplanet transits, eclipsing binaries, or other variable stars, optionally in real time at the telescope.   

\subsection{Single Aperture Photometry}\label{sec:singleaperture}

Single aperture photometry measures the flux from a source within a predefined region of interest in an image referred to as an aperture. AIJ currently supports circular apertures only. A representation of an object's flux in the aperture, referred to as net integrated counts, is calculated by summing all of the pixel values within the aperture after subtracting an estimate of the background flux near the aperture. The background flux is estimated from the pixel values in a background annulus centered on the aperture.

Single aperture photometry can be performed by simply placing the mouse pointer near the center of an object in an image. The net integrated counts within the mouse pointer aperture are shown in the \textit{Int Cnts} display area above the image. The aperture and background regions are set by double clicking the single aperture icon $\left(\,\includegraphics[scale=0.37, trim=1.7mm 1.9mm 1.5mm 0mm]{aijapmodeicon}\,\right)$ on the toolbar, or single clicking the set aperture icon $\left(\,\includegraphics[scale=0.37, trim=1.7mm 2.7mm 1.7mm 0mm]{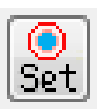}\,\right)$ above an image. The mouse pointer photometer calculates net integrated counts by summing all of the pixel values in the mouse pointer aperture after subtracting the average value of the sky background annulus from each pixel value.

The results of a single aperture measurement can be recorded in a ``measurements table'' (see Appendix \ref{app:measurementstable}) by enabling the single aperture icon $\left(\,\includegraphics[scale=0.37, trim=1.8mm 2.3mm 1.5mm 0mm]{aijapmodeicon}\,\right)$ on the toolbar and then left-clicking near a star. Alternatively, shift-left-click near a star (without enabling single aperture photometry mode) as a short cut for single aperture photometry. These two methods of performing single aperture photometry can optionally use the more sophisticated centroid and sky background measurement algorithms described in \S \ref{sec:multiaperture}.

An azimuthally averaged radial profile of an object can be plotted by left-clicking near the object in an image, and then selecting \textit{Analyze}$\rightarrow$\textit{Plot seeing profile} in the menus above the image display. As a short cut, alt-left-click near an object to produce the plot. If centroid is enabled, the radial profile will be centered on the object. Click the \textit{Save Apertures} button in the plot to use the aperture radii determined from the radial profile. See Appendix \ref{sec:radialprofile} for more details.

AIJ calculates photometric error as described in Appendix \ref{app:photerr}. For proper error calculation, the gain, dark current, and read out noise of the CCD detector used to collect the data must be entered in the \textit{Aperture Photometry Settings} panel (see Appendix \ref{sec:photset}).

\subsection{Multi-Aperture Differential Photometry}\label{sec:multiaperture}

Differential photometry measures the flux of a target star relative to the combined flux of one or more comparison stars. The differential measurement is conducted by performing single aperture photometry on one or more target stars and one or more comparison stars. Then a target star's differential flux is calculated by dividing the target star's net integrated counts, $F_{T}$, by the sum of the net integrated counts of all comparison stars (i.e. the sum of $F_{C_i}$, where $i$ ranges from 1 to the number of comparison stars $n$). The calculation is:
\begin{equation}
\rm{rel\_flux\_T\_j} =  \frac{F_T}{\sum_{i=1}^{n} F_{C_i}},
\label{eq:difffluxT}
\end{equation}
\noindent where $j$ indicates the target star aperture number and $i$ indexes all comparison star aperture numbers. A measurements table data column labeled $\rm{rel\_flux\_T\_j}$ contains the differential flux measurement for target star aperture $j$. The terms relative flux and differential flux are used in this work and in AIJ interchangeably. Differential photometric error is calculated as described in Appendix \ref{app:photerr}. For proper error calculation, the gain, dark current, and read out noise of the CCD detector used to collect the data must be entered in the \textit{Aperture Photometry Settings} panel (see Appendix \ref{sec:photset}).

AIJ also calculates differential flux for each comparison star aperture by comparing the flux in its aperture to the sum of the flux in all \textit{other} comparison star apertures. The calculation is:
\begin{equation}
\rm{rel\_flux\_C\_j} =  \frac{F_{C_j}}{\sum_{i=1}^{n} F_{C_i}, i\ne j},
\label{eq:difffluxC}
\end{equation}
\noindent where $j$ indicates the comparison star aperture number for which differential flux is being calculated and $i$ indexes all comparison star aperture numbers. A measurements table data column labeled $\rm{rel\_flux\_C\_j}$ contains the differential flux measurement for comparison star aperture $j$.

\begin{figure*}
\begin{center}
\resizebox{0.6\textwidth}{!}{
\includegraphics*{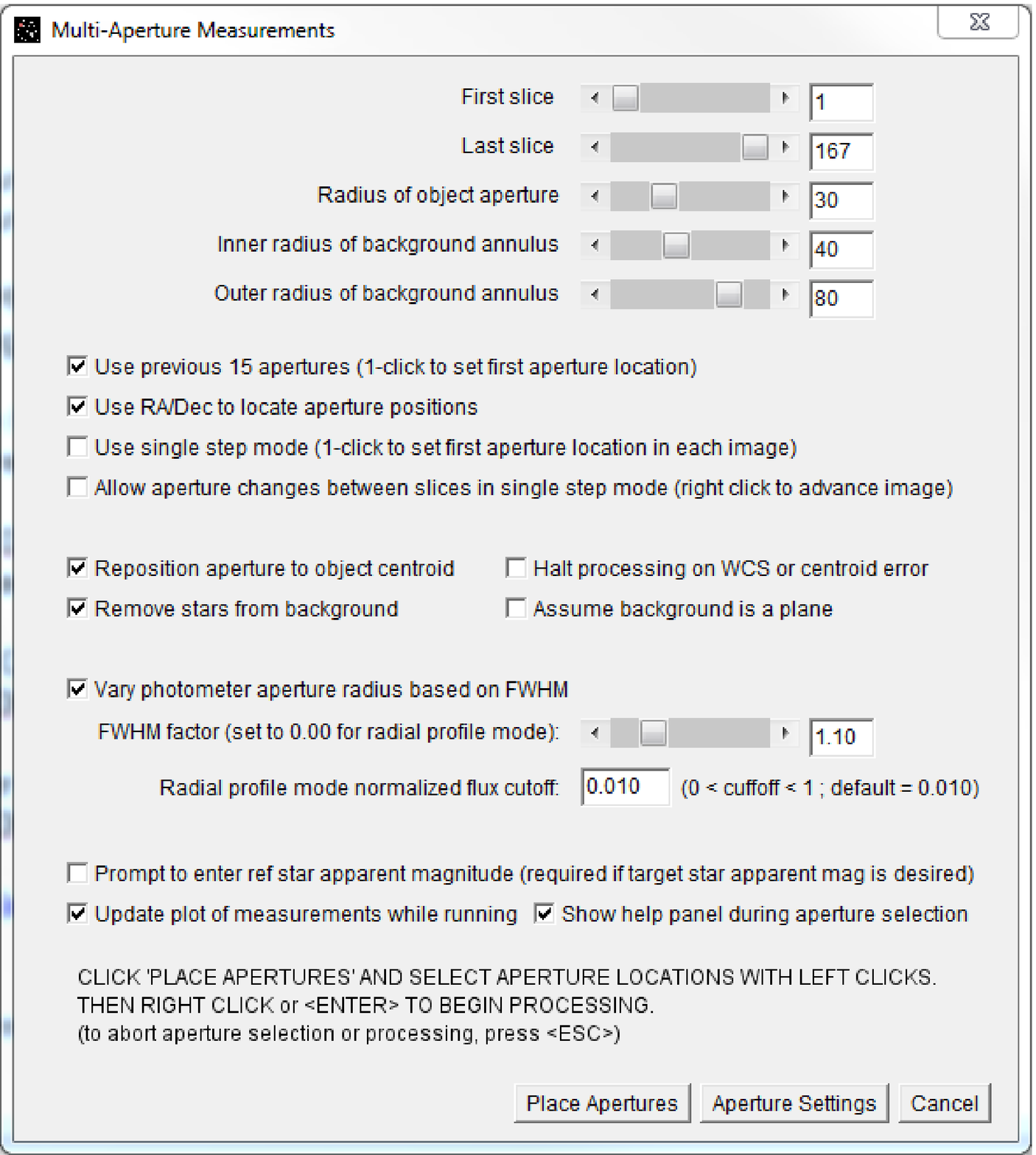} }
\caption{The \textit{Multi-Aperture Measurements} set up panel. Multi-Aperture automates the task of performing differential photometry on a time-series of images. Various settings are available in the set-up panel, and then the apertures are placed and adjusted interactively by clicking near stars directly in the image display. See text for more details.}
\label{fig:aijmultiaperture}
\end{center}
\end{figure*}

Multi-Aperture (MA) automates the task of performing differential photometry on a time-series of images. Various settings are presented in a set-up panel, and then the target and comparison star apertures are placed and adjusted interactively by clicking near stars directly in the image display. The MA module is launched by clicking the MA icon $\left(\,\includegraphics[scale=0.37, trim=1.5mm 2.4mm 1.6mm 0mm]{aijmultiapertureicon}\,\right)$ above an image or in the AIJ Toolbar. The \textit{Multi-Aperture Measurements} set-up panel shown in Figure \ref{fig:aijmultiaperture} opens. The top two scroll bars allow the user to set the range of  images to be processed. The three scroll bars immediately below allow the user set the aperture radius, the inner radius of the sky background region, and the outer radius of the sky background region, all in units of pixels.

The \textit{Use previous apertures...} option allows the previously defined set of apertures to be re-used. Aperture definitions can also be stored and reopened from disk in the \textit{File} menu of the image display. If WCS headers are available and the \textit{Use RA/Dec...} option is selected, the saved apertures will be placed 
according to RA and Dec rather than by $x$ and $y$ pixel coordinates. If the \textit{Reposition aperture to object centroid} option is selected, a centroid algorithm will attempt to center the aperture on the nearest star.

The \textit{Remove stars from background} option enables iterative $2\sigma$ cleaning of the sky-background region. The iteration continues until the mean ADU of the pixels remaining in the background set converges or the maximum number of
allowed iterations has been reached without convergence. The sky background
pixels remaining after the cleaning operation are used to calculate and remove
the sky background at each pixel in the aperture. If the \textit{Assume background is a plane} option is selected, AIJ fits a plane to the remaining pixels in the background region and subtracts the value of the plane at each pixel within the aperture to remove the sky background contribution. Otherwise, AIJ subtracts the mean of the remaining pixels in the background annulus from each pixel in the  aperture.

If the \textit{Vary photometer aperture radius based on FWHM} option is enabled, the aperture radius used in an individual image of a time-series is equal to the product of the user specified \textit{FWHM factor} and the average full width at half maximum (FWHM) from all apertures in that image. This mode may improve photometric precision when seeing varies significantly or telescope focus drifts within a time-series. If the \textit{FWHM factor} is set to ``0.00'', an azimuthally averaged radial profile (see Appendix \ref{sec:radialprofile}), centered on the aperture, is used to determine the aperture radius based on the \textit{Radial profile mode normalized flux cutoff} value specified. In this mode, the aperture radius used in an image is equal to the distance from the center of the aperture at which the radial profile value is equal to the specified normalized flux cutoff. The variable aperture modes should not be used in crowded fields since the changing aperture radius will worsen the effects of variable amounts of contaminating flux blending into the aperture as seeing changes. 

If the \textit{Prompt to enter ref star apparent magnitude} option is enabled, the apparent magnitude of target aperture sources will be calculated from user entered apparent magnitudes of one or more comparison aperture sources. The target aperture apparent magnitude data can be formatted to submit to the Minor Planet Center using the \textit{Multi-plot Main} panel menu item \textit{File}$\rightarrow$\textit{Create Minor Planet Center format}. See Appendix \ref{app:apparentmagnitude} for details on the use of the apparent magnitude feature. 

The \textit{Aperture Settings} button at the bottom provides access to two panels containing detailed settings for the photometric measurements (see Appendix \ref{sec:photset}). When all options have been set, the \textit{Place Apertures} button causes the set-up panel to close and the program waits for the aperture positions to be identified by user clicks near stars in the first image of the time-series. If the \textit{Use previous apertures...} option is enabled, the user clicks near the first star in the set (usually the primary target star), and all other apertures are placed relative to the first aperture. If the \textit{Use RA/Dec...} option is also enabled, no clicks are required to apply the stored apertures since they are placed automatically at the calculated pixel locations corresponding to the stored WCS coordinates. If photometry settings (e.g. aperture radii) need to be refined during aperture placement, click the set aperture icon $\left(\,\includegraphics[scale=0.38, trim=1.8mm 2.8mm 2mm 0.2mm]{aijapseticon}\,\right)$ above the image display.

By default, a help panel opens that describes actions available at each step during aperture placement. By default, the first left-click in an image places a target star aperture and all other clicks place comparison star apertures. Using shift-left-click reverses the sense of the default aperture type when placing new apertures. A shift-left-click in an existing aperture changes it from target to comparison, and vice-verse. A left-click inside an existing aperture deletes the aperture. An aperture can be moved by left-clicking inside it and dragging it to a new position. Other options are available as listed in the context sensitive help panel.

When all apertures have been defined as desired, a right-click or press of the \textit{$<$Enter$>$} key will start the automated differential photometry process on all images defined in the set-up panel. If the \textit{Use RA/Dec...} option is enabled, apertures will first be positioned in each image of the series according to the WCS information in the image header. If not, the aperture placements start on each subsequent image in the series at the same place in $x,y$ space as in the previous image. In both cases, if centroid is enabled for a particular aperture, the centroid function will attempt to center the aperture on the nearest star. For telescopes with poor tracking/guiding or requiring a meridian flip during the time-series, the images should be plate solved so that the \textit{Use RA/Dec...} option can be enabled to properly find the initial aperture placements in each image. 

The centroid function can be enabled or disabled on a per aperture basis. An alt-left-click in an existing aperture inverts the sense of centroid for that aperture. An aperture with centroid enabled has a plus sign showing at its center. If the first aperture has the centroid function enabled, all non-centroided apertures will move from one image to the next based on the average movement of all centroided apertures. This allows apertures to be placed around faint stars that are near bright stars that would normally capture the aperture if centroid were enabled. 

The photometry data is written to a ``measurements table'' (see Appendix \ref{app:measurementstable}) and optionally plotted by the Multi-Plot program as MA progresses through the time-series. If MA is run a second time duing the same AIJ session, the new measurements will be appended to the existing measurements in the measurements table. To clear the measurements table before the second run, click the clear measurements icon $\left(\,\includegraphics[scale=0.38, trim=1.7mm 2.5mm 1.7mm 0mm]{aijcleartableicon}\,\right)$ above the image display.

\subsection{Multi-Plot}\label{sec:multiplot}

Multi-Plot (MP) provides a multi-curve plotting facility that is tightly integrated with differential photometry and light curve fitting. MA can automatically start MP, or MP can be manually started by clicking the MP icon $\left(\,\includegraphics[scale=0.5, trim=0.9mm 1.85mm 1.1mm 0mm]{aijmultiploticon}\,\right)$ on the AIJ Toolbar. If a measurements table has been created by MA or opened from the OS, MP will automatically create a plot based on the last plot settings. Alternatively, plot templates can be saved and restored to easily format commonly created plots. Plotting controls are accessed in two main user interface panels.

\begin{figure*}
\begin{center}
\resizebox{\textwidth}{!}{
\includegraphics*{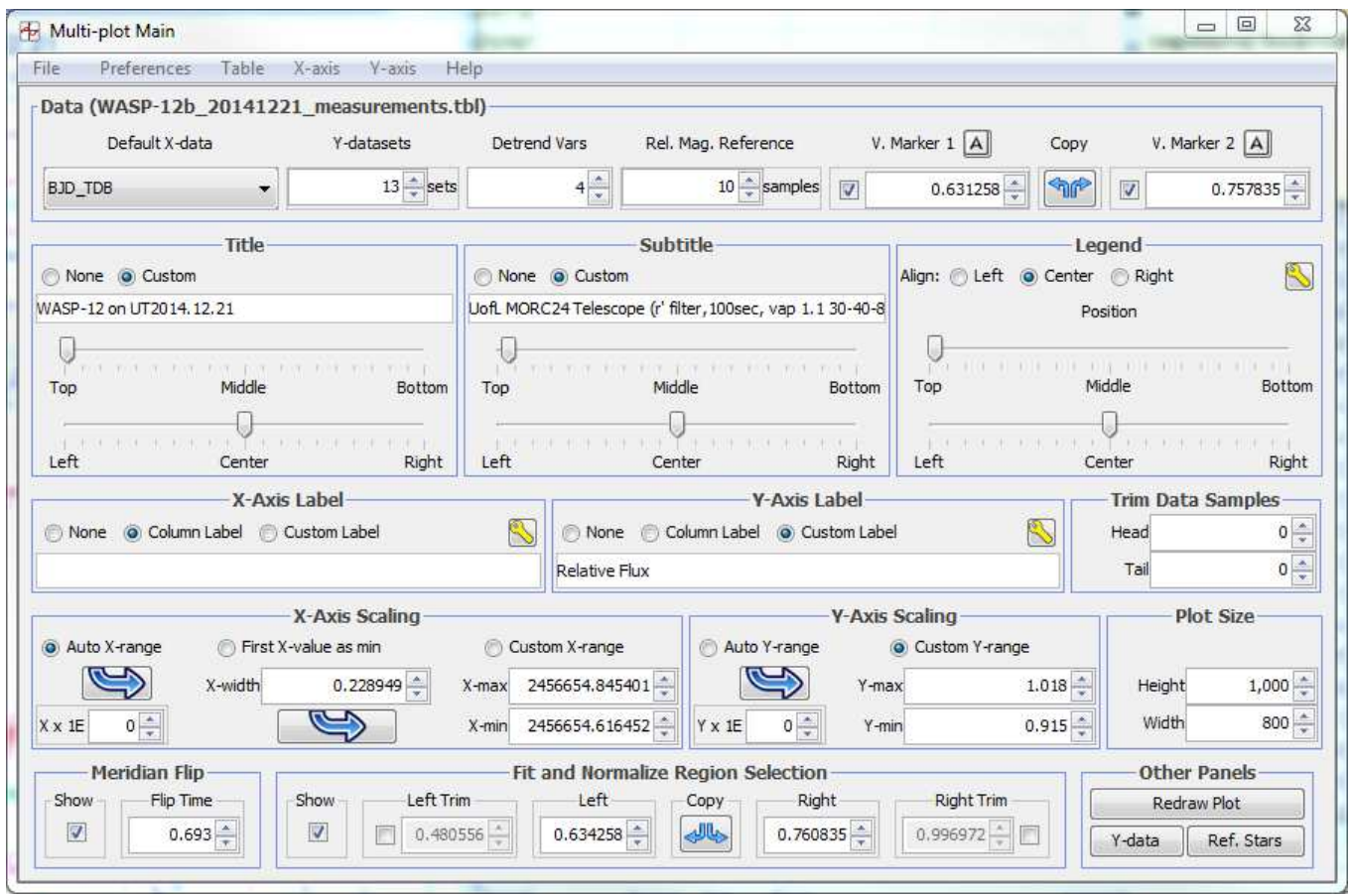} }
\caption{The AIJ \textit{Multi-plot Main} panel. The \textit{Multi-plot Main} panel provides access to plotting controls that affect the overall plot. Important controls include the default $x$-axis dataset (usually a time dataset such as $\bjdtdb$), the title, legend, axis labels,  plot size, and axis scaling settings. The bottom row provides several settings used for detrending, normalization, and light curve modeling. See text for more details.}
\label{fig:aijmpmain}
\end{center}
\end{figure*}

\begin{figure*}
\begin{center}
\resizebox{\textwidth}{!}{
\includegraphics*{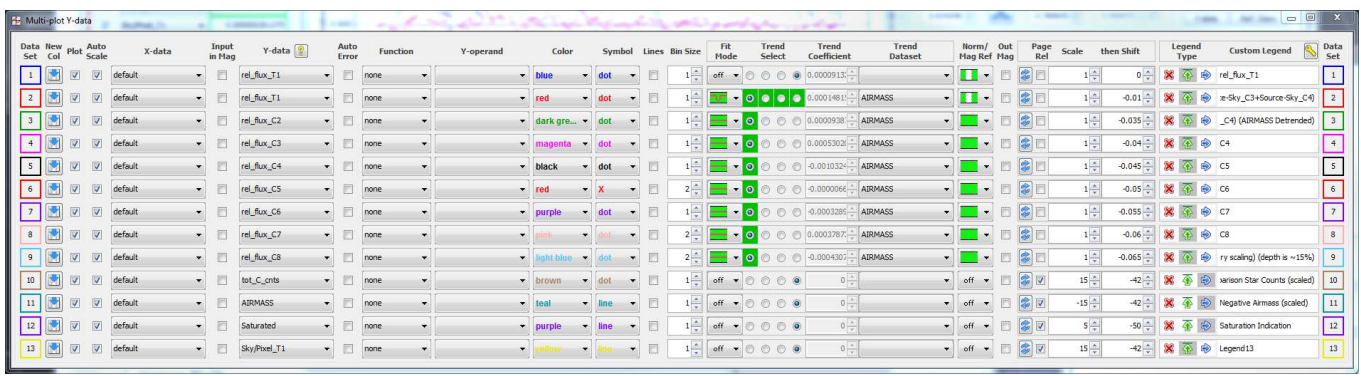} }
\caption{The AIJ \textit{Multi-plot Y-data} panel. The \textit{Multi-plot Y-data} panel provides access to plotting controls that affect individual plotted datasets. Each horizontal row provides controls for plotting the dataset selected under the \textit{Y-data} heading. See text for more details.}
\label{fig:aijmpydata}
\end{center}
\end{figure*}

The \textit{Multi-plot Main} panel is shown in Figure \ref{fig:aijmpmain}. Controls include selection of the default $x$-axis dataset from a pull-down list of all data columns in the measurements table, the maximum number of plotted datasets and the maximum number of detrend variables displayed in the \textit{Multi-plot Y-data} panel shown in Figure \ref{fig:aijmpydata}, plot title and subtitle, legend placement and options, $x-$ and $y-$axis label and scaling options, and the overall plot size in pixels. The \textit{V.Marker 1} and \textit{V.Marker 2} controls provide the option to plot up to two vertical red dashed lines with labels. For example, these lines can be set to mark the predicted ingress and egress times on exoplanet transit light curve plots. As discussed in the last paragraph of Appendix \ref{sec:coordinateconverter}, new astronomical data such as AIRMASS, $\bjdtdb$, etc. can be calculated and added to the measurements table using the menu item at \textit{Multi-plot Main}$\rightarrow$\textit{Table}$\rightarrow$\textit{Add new astronomical data columns to table}. 

The bottom row of \textit{Multi-plot Main} sub-panels provide access to other $x$-axis controls that define the regions used when normalizing, detrending, and fitting $y$-datasets. The \textit{Meridian Flip} settings allow the time of the telescope's meridian flip to be specified (if applicable) and to optionally mark that time with a light blue dashed vertical line. If \textit{meridian\_flip} is selected as a detrending parameter for a dataset in the \textit{Multi-plot Y-data} panel, the fitting routine attempts to remove any baseline offsets from one side of the meridian flip to the other. The \textit{Fit and Normalize Region Selection} options allow the user to specify additional $x$-axis values that define the normalization, detrend, and fit regions. For example, the \textit{Left} and \textit{Right} settings are used to mark the regions used to normalize the data (which normally exclude the in-transit portion of the light curve). Also, the $x$-axis mid-point between the \textit{Left} and \textit{Right} settings is used as the model fit starting point for the transit model parameter $T_C$. The \textit{Left Trim} and \textit{Right Trim} settings can be used to exclude leading and/or trailing data from the normalization, detrending, and fitting processes. A gray vertical dashed line can optionally be displayed to identify the regions on the plot. Additional options are available in the menus at the top of the \textit{Multi-plot Main} panel.

\begin{figure*}
\begin{center}
\resizebox{0.85\textwidth}{!}{
\includegraphics*{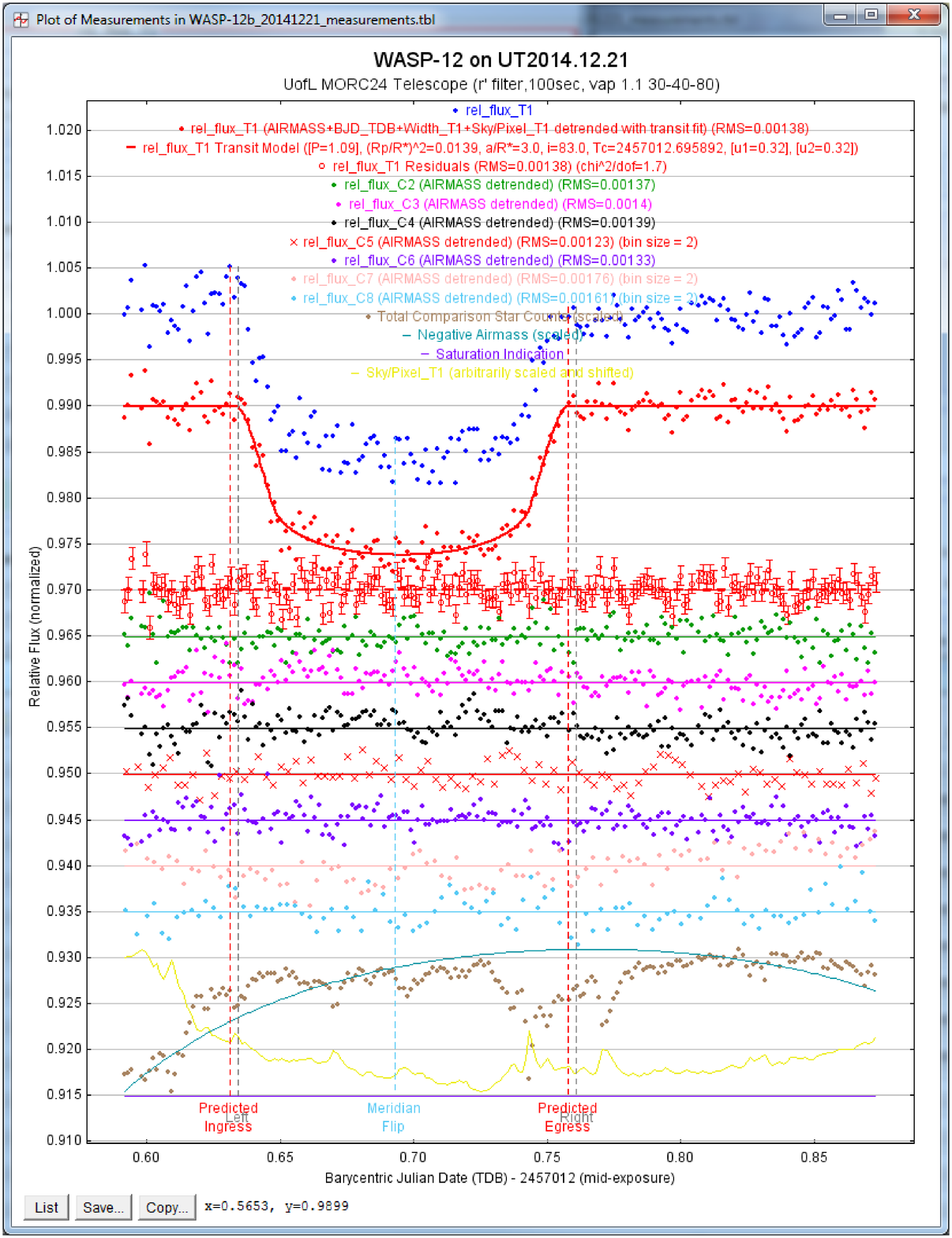} }
\caption{Multi-Plot example plot. The plot of a WASP-12b transit under poor observing conditions is shown. The top dataset plotted with solid blue dots is the raw normalized differential photometry. The solid red dots show the light curve after simultaneously detrending and fitting with an exoplanet transit model, which is shown by the red line through the data. Note the reduced systematics and scatter in the detrended data. The open red dots show the model residuals with error bars. See the text for descriptions of the other plotted data.}
\label{fig:aijmpplot}
\end{center}
\end{figure*}

Figure \ref{fig:aijmpplot} shows an example plot of a WASP-12b transit and demonstrates many of MP's plotting capabilities. The settings shown in Figures \ref{fig:aijmpmain} (\textit{Multi-plot Main}), \ref{fig:aijmpydata} (\textit{Multi-plot Y-data}), and \ref{fig:aijfitpanel} (\textit{Data Set 2 Fit Settings}) were used to produce this plot. 

The \textit{Multi-plot Y-data} panel is shown in Figure \ref{fig:aijmpydata}. Each horizontal row in the user interface corresponds to an individual plotted dataset. The example shown allows up to 13 datasets to be plotted on a single plot. The top row of controls labeled ``1'' under the \textit{Dataset} heading produce the raw normalized light curve shown as solid blue dots near the top of the plot. Note in particular the \textit{Y-data} column selected (rel\_flux\_T1) and the \textit{Norm/Mag Ref} mode selected $\left(\,\includegraphics[scale=0.4, trim=1.4mm 2.1mm 1.3mm 0mm]{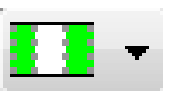}\,\right)$. The green areas in the \textit{Norm/Mag Ref} mode icon indicate the regions of the light curve (relative to the \textit{Left} and \textit{Right} markers on the \textit{Multi-plot Main} panel) that are used to normalize the data. In this case, the in-transit data are not included in the calculation of the normalization parameter. 

The second row of plot controls (\textit{Dataset 2}) again plot the rel\_flux\_T1 data, but this time after simultaneously detrending and fitting the data as set up in the fit panel shown in Figure \ref{fig:aijfitpanel}, which is discussed in \S \ref{sec:fitanddetrend}.  Note the reduced systematics and scatter in the data. This row of controls uses the \textit{Fit Mode selection} $\left(\,\includegraphics[scale=0.4, trim=1.3mm 2.1mm 1.3mm 0mm]{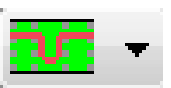}\,\right)$ to enable the fit panel. Note that the \textit{then Shift} column is set to $-0.01$ which shifts the transit baseline down to $y=0.990$ on the plot for clarify. The light curve model residuals are shown as open red circles. The residuals plot controls are available in the fit panel corresponding to the light curve.

The plot controls on rows corresponding to \textit{Dataset 3} through \textit{Dataset 9} display the first seven comparison star differential light curves on the plot. The \textit{Fit Mode} selection $\left(\,\includegraphics[scale=0.4, trim=1.2mm 2.1mm 1.4mm 1mm]{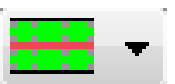}\,\right)$, being completely green, selects all data for detrending, and the flat red line indicates that no transit model is fit (since no transit event is expected in the comparison stars). Note that the normalize mode selected under the \textit{Norm/Mag Ref} heading is all green also, since all comparison star data can be used for normalization. Each comparison star light curve is shifted from the other light curves for clarity using the \textit{then Shift} setting. Note that datasets 5, 8, and 9 have been binned by 2 data samples using the \textit{Bin Size} setting to reduce the scatter for plotting purposes. The current implementation of binning is actually averaging of the specified number of data points, rather than binning into fixed size $x$-axis bins.

Each comparison star light curve has been detrended against airmass. In this example, MP is set to allow up to four trend datasets to be selected for detrending. Note that the first \textit{Trend Select} button is enabled and \textit{AIRMASS} is showing in the \textit{Trend Dataset} column for all comparison stars. If the resulting comparison star light curves are relatively flat, then they should perform well as part of the comparison ensemble.

The plot controls for datasets 10-13 create the display of four diagnostic curves. These curves are plotted relative to the size of the plot page by selecting the \textit{Page Rel} option on each row. In this mode, the \textit{Scale} setting forces the plot of the data to fit within a fixed percentage of the plot range. The \textit{then Shift} value in this mode is also a percent of the plot range with 0 being in the middle of the plot. This mode makes scaling of data to fit on a plot easy when the shape of a curve is important, but the actual values of the data are not.

Many other plotting options are available, including plotting of error bars (\textit{Auto Error}), legend options, input and output in magnitudes, data averaging/binning, and the option to select independent $x$-axis datasets for each $y$-axis dataset (i.e. the \textit{X-data} pull-down menus in the \textit{Multi-plot Y-data} panel). 

Datasets displayed in a plot have typically been modified in one or more ways (e.g. normalized, detrended, converted to/from magnitude, scaled, shifted, binned, etc.). The displayed values can be added to the measurements table as new data columns for further manipulation or permanent storage using the \textit{New Col} button $\left(\,\includegraphics[scale=0.39, trim=1.3mm 2.4mm 1.2mm 0mm]{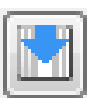}\,\right)$ on the left-hand side of the row corresponding to the plotted data. Model residuals and sampled versions of the model can also be saved to the measurements table using this button.

The legend is shown at the top of the plot in Figure \ref{fig:aijmpplot}, with the light curves plotted below. Legend entries for each plotted curve can be automatically generated based on the measurements table column names known to be produced by MA by selecting the \textit{Legend Type} icon $\left(\,\includegraphics[scale=0.38, trim=0.9mm 2.2mm 1.1mm 0mm]{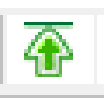}\,\right)$. A custom legend can be displayed individually or in combination with the automatic legend by enabling the \textit{Legend Type} icon $\left(\,\includegraphics[scale=0.35, trim=1mm 2.7mm 1.2mm 0mm]{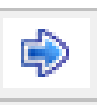}\,\right)$, or a curve's legend can be disabled by selecting the $\left(\,\includegraphics[scale=0.35, trim=1.5mm 2.4mm 1.0mm 0mm]{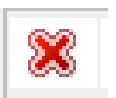}\,\right)$ option. Legend settings that affect all legend entries are accessed by clicking the \textit{Legend Preferences} icon $\left(\,\includegraphics[scale=0.4, trim=1.6mm 2mm 1.5mm 0mm]{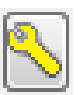}\,\right)$ at the top-right of the \textit{Multi-plot Y-data} panel. By default, the model residuals RMS values are calculated and displayed in the legend for detrended and fitted light curves. The light curve model parameter values are optionally shown for fitted light curves. The predicted time of ingress and egress are shown as red dashed vertical lines. The meridian flip time is indicated by the light blue vertical dashed line (although no meridian flip actually occurred for the example observations), and the \textit{Left} and \textit{Right} gray dashed vertical lines show the boundaries of the normalization, detrending, and fitting regions.

The plot can be zoomed by placing the mouse pointer inside the plot image and rolling the mouse scroll wheel. Also, a left-click in the plot image zooms in by one step, and a right-click zooms out to the full plot. After zooming in, the plot can be panned with a left-click and drag of the mouse. A data point can be removed from a plot and measurements table by holding the shift key and moving the mouse over the data point in the plot until it is highlighted. Then, while continuing to hold shift, a left-click will remove the data point. A shift-right-click will retrieve the deleted point. As shortcuts, the \textit{Left}, \textit{Right}, \textit{Left Trim}, and \textit{Right Trim} vertical markers can be set at the mouse pointer location in the plot with a control-left-click,  control-right-click, control-shift-left-click, and control-shift-right-click, respectively.

All typical data and image products created by a photometry and/or plotting and fitting session can be saved with one action using the \textit{Save All} feature available in the \textit{File} menu of the \textit{Multi-plot Main} panel. There are also \textit{File} menu options in \textit{Multi-plot Main} and image display panels to save each data product separately. The menu option at \textit{File}$\rightarrow$\textit{Save all (with options)} opens a \textit{Save All} settings panel before saving the files. The menu option at \textit{File}$\rightarrow$\textit{Save all} runs the process using the previous settings without opening the settings panel. See Appendix \ref{sec:saveall} for a full description of the \textit{Save All} feature.

\subsection{Light Curve Fitting and Detrending}\label{sec:fitanddetrend}

\begin{figure*}
\begin{center}
\resizebox{\textwidth}{!}{
\includegraphics*{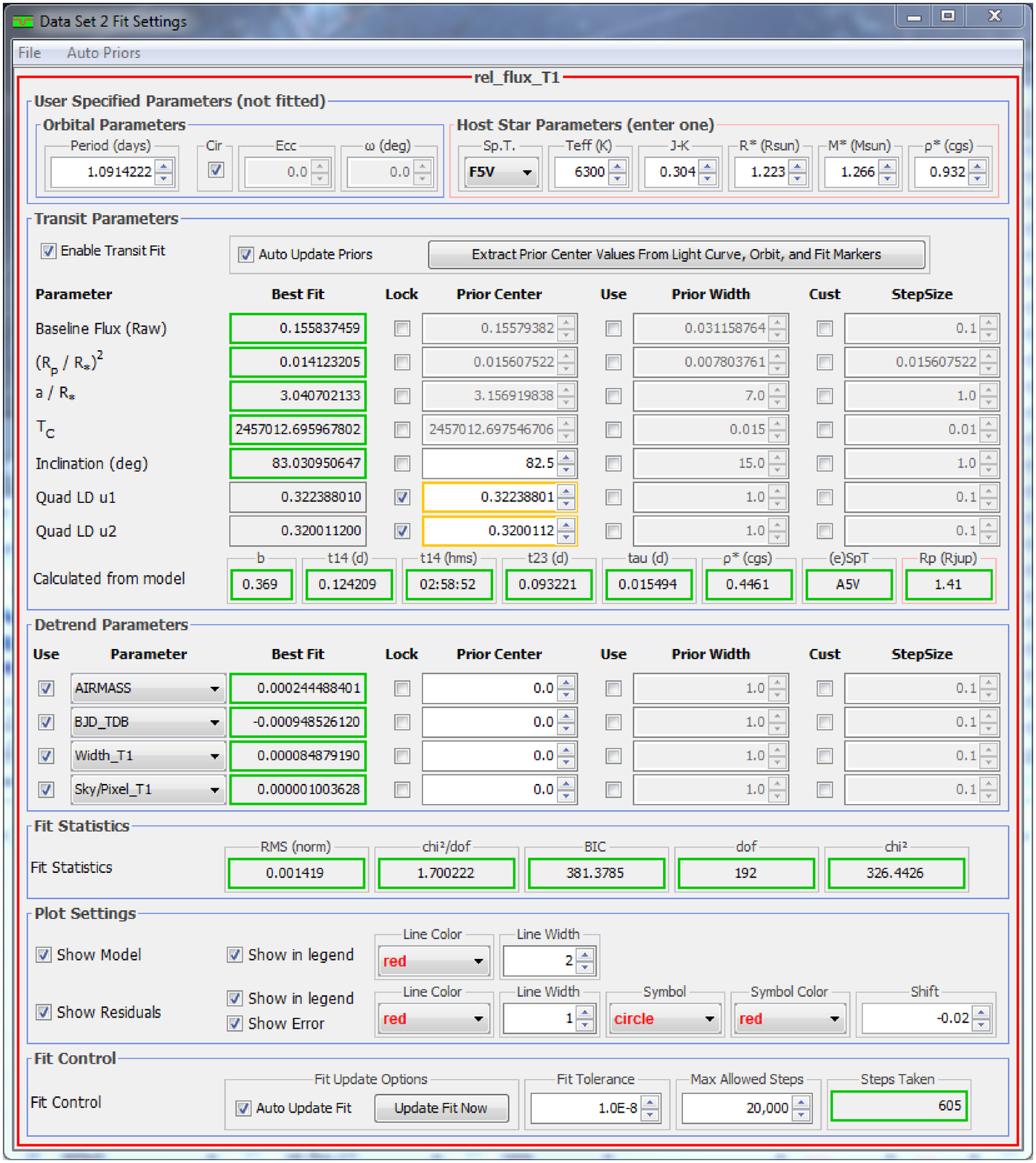} }
\caption{The \textit{Fit Settings} panel. The light curve model fitting and detrending panel settings shown here produce the light curve model fit in the example plot of Figure \ref{fig:aijmpplot}. Light curve prior center and width values can optionally be set by the user to constrain the model fit. Direct access to all detrending parameters is also provided along with more flexible settings. Several values calculated from the model parameters are displayed, along with several statistical values useful for assessing data quality and goodness of the model fit to the data. See text for more details.}
\label{fig:aijfitpanel}
\end{center}
\end{figure*}

Light curve fitting is enabled for a particular dataset in the \textit{Multi-plot Y-data} panel by selecting the \textit{Fit Mode} icon showing the red transit model on a full green background $\left(\,\includegraphics[scale=0.4, trim=1.3mm 2.1mm 1.3mm 0mm]{aijmpfiticon}\,\right)$. When this mode is selected, a \textit{Fit Settings} panel will be displayed for the dataset as shown in Figure \ref{fig:aijfitpanel}. The settings in the figure produce the light curve model fit shown in the example plot of Figure \ref{fig:aijmpplot}. The transiting exoplanet model is described in \citet{Mandel:2002}. The transit is modeled as an eclipse of a spherical star by an opaque planetary sphere. The model is parametrized by six physical values, plus a baseline flux level, $F_0$. The six physical parameters are the planetary radius in units of the stellar radius, $R_P/R_*$, the semi-major axis of the planetary orbit in units of the stellar radius, $a/R_*$, the transit center time, $T_C$, the impact parameter of the transit, $b$, and the quadratic limb darkening parameters, $u_1$ and $u_2$. The orbital inclination can be calculated from the model parameters as
\begin{equation}
i=\cos^{-1}\left(b\frac{R_*}{a}\right).
\end{equation} 

AIJ is currently limited to finding the best fit model parameter values and does not provide estimates of the parameter uncertainties. The best fit model is found by minimizing $\chisq$ of the model residuals using the downhill simplex method to find local minima \citep{Nelder:1965}.

Before making adjustments in the \textit{Fit Settings} panel, refer \S \ref{sec:multiplot} and properly configure the \textit{Fit and Normalize Region Selection} settings and optionally the \textit{Meridian Flip} settings in the \textit{Multi-plot Main} panel, and the \textit{Norm/Mag Ref} mode and other settings in the corresponding row of the \textit{Multi-plot Y-data} panel.

The parameter settings in the \textit{User Specified Parameters} sub-panel are not fitted. The period of the exoplanet's orbit is not well constrained by the transit model, but its value will impact the best fit values of some of the fitted parameters, so the \textit{Period} value must be entered by the user. Normally, the orbital period is known from RV or wide-field survey photometric data. The \textit{Host Star Parameters} in the same sub-panel are only used to estimate the physical planet radius, $R_P$, from the fitted parameter $R_P/R_*$. The host star parameter values are interrelated by tables in \citet{Allen:2001} for zero age main sequence (ZAMS) stars. The only value used in the calculation of $R_P$ (displayed near middle of the right-hand side of the panel) is $R_*$, so that value should be entered directly if known. Otherwise, entering any one of the other host star parameters will produce a rough estimate of $R_*$ based on the ZAMS assumption.

The \textit{Transit Parameters} sub-panel has seven rows for the seven transit model parameters. \textit{Prior Center} values will need to be set for the seven parameters to ensure the correct $\chisq$ minimum is found. The top four parameters shown in the sub-panel are extracted from the light curve data by default. In the odd case that those estimated values are not correct, the values can be entered directly by the user. In the example shown, the \textit{Inclination} prior center value has been set by the user, but no constraints have been placed on the range of valid final fitted values (although the upper end is limited to $90\degrees$ by the definition of inclination). The \textit{Quad LD u1} and \textit{Quad LD u2} parameter values have been set by the user, and the fitted values have been locked to those values by enabling the \textit{Lock} option beside each one. The fixed values of u1 and u2 were extracted from the \citet{Claret:2011} theoretical models using a website tool\footnote{http://astroutils.astronomy.ohio-state.edu/exofast/limbdark.shtml}. The best fit transit model parameter values are displayed in the \textit{Best Fit} column. A green box around the fitted parameter values indicates that the minimization converged to a value less than the \textit{Fit Tolerance} within the \textit{Max Allowed Steps}. Both of those minimization parameters can be set at the bottom of the fit panel in the \textit{Fit Control} sub-panel, but the default values normally work well.

The bottom row in the \textit{Transit Parameters} sub-panel shows several values that are calculated from the best fit model. The host star's density, $\rho_*$, is particularly interesting, since a good estimate can be derived from the transit light curve data alone. Positioning the mouse pointer over any parameter will optionally cause a description of the parameter to be temporarily displayed.

The \textit{Prior Width} column allows the user to optionally limit the range of a parameter's fitted value. \textit{Prior Width} values are not normally needed, but may be helpful in fitting an ingress- or egress-only partial transit. The \textit{StepSize} column allows the user to set a custom initial minimization step size. However, the default values for each parameter normally work well, so setting custom values is not usually necessary.

The \textit{Detrend Parameters} sub-panel duplicates the detrend settings on the \textit{Multi-plot Y-data} panel. However, the \textit{Fit Settings} panel provides direct access to all detrend parameters and settings. Prior center values, widths, and fitting step sizes can optionally be set for detrend parameters as well. 

Light curve detrending is accomplished by including a $\chisq$ contribution for each selected detrend parameter in the overall light curve fit. The $\chisq$ contribution at each step of the minimization represents the goodness of the linear fit of the detrend parameters to the light curve after subtracting the light curve model corresponding to the current fit step. The $\chisq$ contribution for all $n$ detrend parameters is calculated at each step of the fitting process as
\begin{equation}
\chisq_{D} = \sum_{k=1}^{m} \frac{\left(O_k-\left(\sum_{j=1}^{n}c_{j}D_{j_k}\right)-E_k\right)^2}{\sigma_k^2},
\label{eq:chi2fordetrend}
\end{equation}
\noindent where $j$ indexes the detrend parameters, $k$ indexes the samples of the light curve, $m$ is the total number of samples in the light curve, $O_k$ is the observed normalized differential target flux, $c_j$ is the fitted linear coefficient for the detrend parameter values $D_{j_k}$, $E_k$ is the expected value of the flux (which is the normalized transit model value corresponding to the time of the $k^{\rm th}$ data sample), and $\sigma_k$ is the error in the normalized differential target flux for each sample.

The \textit{Fit Statistics} sub-panel lists five statistical values that allow the user to assess the quality of the data and goodness of the model fit to the data. The values displayed from left to right are RMS of the model residuals, $\chisq$ per degree of freedom (i.e. reduced $\chisq$), Bayesian Information Criterion (BIC), the number of degrees of freedom, and the total $\chisq$. BIC is defined as
\begin{equation}
{\rm{BIC}}=\chisq+p\ln n,
\end{equation}
\noindent where $p$ is the number of fitted parameters, and $n$ is the number of fitted data points. The BIC can be used to determine whether the addition of a new parameter to a model (in particular an optional one such as a detrend parameter) provides a significant improvement in the fit. If the BIC value decreases by more than 2.0 when a model parameter is added, then the new model is preferred over the model with fewer parameters. A larger decrease in the BIC value suggests a stronger preference for the new model.

The \textit{Plot Settings} sub-panel provides options for plotting the light curve model and the residuals. The \textit{Fit Control} sub-panel settings are not normally needed, since by default the model fit recalculates any time a value in the panel is changed, and since the default \textit{Fit Tolerance} and \textit{Max Allowed Steps} settings work for most light curve datasets.

\subsection{Comparison Ensemble Management}\label{sec:ensemblemanagement}

\begin{figure}
\begin{center}
\resizebox{\columnwidth}{!}{
\includegraphics*{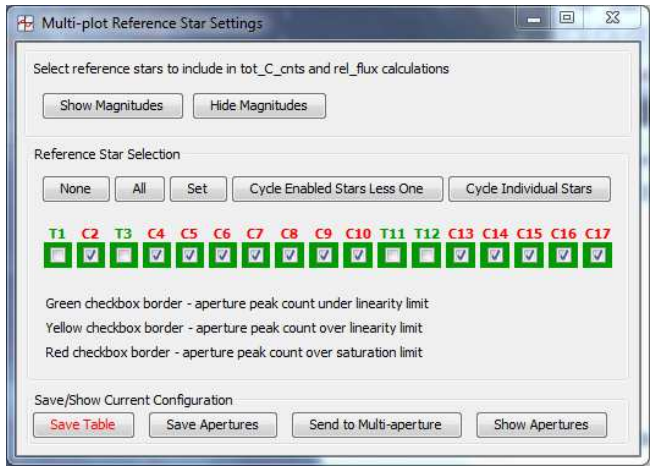} }
\caption{The \textit{Multi-plot Reference Star Settings} panel. Star apertures can be changed from comparison to target apertures and vice-versa. After any change, all relative photometry values are recalculated and the measurements table and plot are updated. Target aperture IDs start with a green T prefix and have a deselected checkbox. Comparison aperture IDs start with a red C prefix and have a selected checkbox. The \textit{Cycle Enabled Stars Less One} and the \textit{Cycle Individual Stars} buttons speed the task of finding and excluding poor comparison stars. See text for more details.}
\label{fig:aijmprefstarpanel}
\end{center}
\end{figure}

The Multi-Plot environment allows the user to include or exclude comparison stars from the comparison ensemble without re-running Multi-Aperture, as long as apertures were defined for all potentially good comparison stars in the original Multi-Aperture run. The \textit{Multi-plot Reference Star Settings} panel shown in Figure \ref{fig:aijmprefstarpanel} provides a checkbox corresponding to each target and comparison star included in the original Multi-Aperture differential photometry run. Deselected stars are considered target stars and selected stars are comparison stars belonging to the comparison ensemble. When a star is added to or removed from the ensemble, the relative flux values for each star are recalculated and the measurements table and plot are updated. 

The \textit{Cycle Enabled Stars Less One} button allows the user to quickly cycle through the comparison ensemble removing one star at a time so that poor comparison stars can be quickly identified and removed from the ensemble. The \textit{Cycle Individual Stars} button can be used to quickly assess the quality of each comparison star individually by cycling through each one as a single comparison star.

\section{AstroImageJ Updater}

\begin{figure}
\begin{center}
\resizebox{\columnwidth}{!}{
\includegraphics*{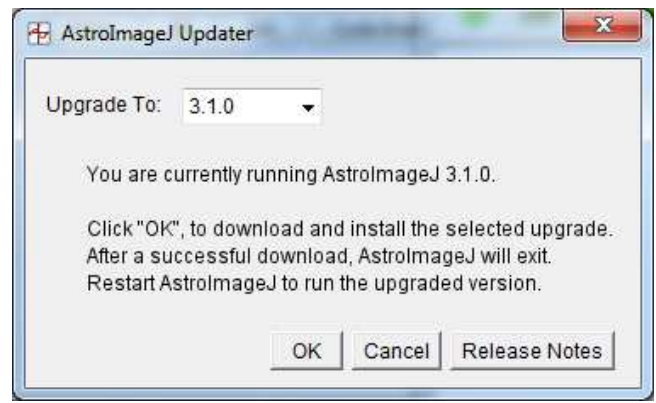} }
\caption{The \textit{AstroImageJ Updater} panel. AIJ updates are easily installed by selecting the desired release number in the \textit{Upgrade To} field. The latest release notes are displayed by clicking the \textit{Release Notes} button. The selected update is installed by clicking the \textit{OK} button. When the installation is finished, AIJ automatically closes. The new version is activated when AIJ is restarted.}
\label{fig:aijupdatepanel}
\end{center}
\end{figure}

AIJ can be easily upgraded to the latest version using the update facility at \textit{AIJ Toolbar}$\rightarrow$\textit{Help}$\rightarrow$\textit{Update AstroImageJ}. The \textit{AstroImageJ Updater} panel shown in Figure \ref{fig:aijupdatepanel} opens. The latest release notes are displayed by clicking the \textit{Release Notes} button. The default value displayed in the \textit{Upgrade To} field is the latest version. The update is installed by clicking the \textit{OK} button. When the installation is finished, AIJ automatically closes. The new version is activated when AIJ is restarted.

\acknowledgments
K.A.C. acknowledges support from NASA Kentucky Space Grant Consortium Graduate Fellowships. 
K.A.C. and K.G.S. acknowledge support from NSF PAARE grant AST-1358862 and the Vanderbilt Initiative in Data-intensive Astrophysics.  
We thank the anonymous referee for a thoughtful reading of the manuscript and for useful suggestions. 
This work has made use of the SIMBAD database operated at CDS, Strasbourg, France.

\appendix

\section{AIJ Utilities}\label{app:utilities}

\subsection{Coordinate Converter}\label{sec:coordinateconverter}

The Coordinate Converter (CC) module converts astronomical coordinates and times to other formats based on observatory location and target coordinates, and integrates AstroImageJ with Simbad and sky-map.org web services. CC can be operated as a module under full control of the user, and it can be operated under the control of DP and MP to provide automated calculations within those modules. When operated by the user, all fields are available to be set as desired. When controlled by DP and MP, only a subset of fields are enabled for user entry, while the other fields are under the control of the program and disabled (grayed-out) to prevent user input. A cross-platform Java based version of CC called AstroCC runs completely independent of AIJ and is available for download from the AIJ website. Detailed CC help is available in the menus above the CC panel at \textit{Help}$\rightarrow$\textit{Help}.

\begin{figure*}
\begin{center}
\resizebox{\textwidth}{!}{
\includegraphics*{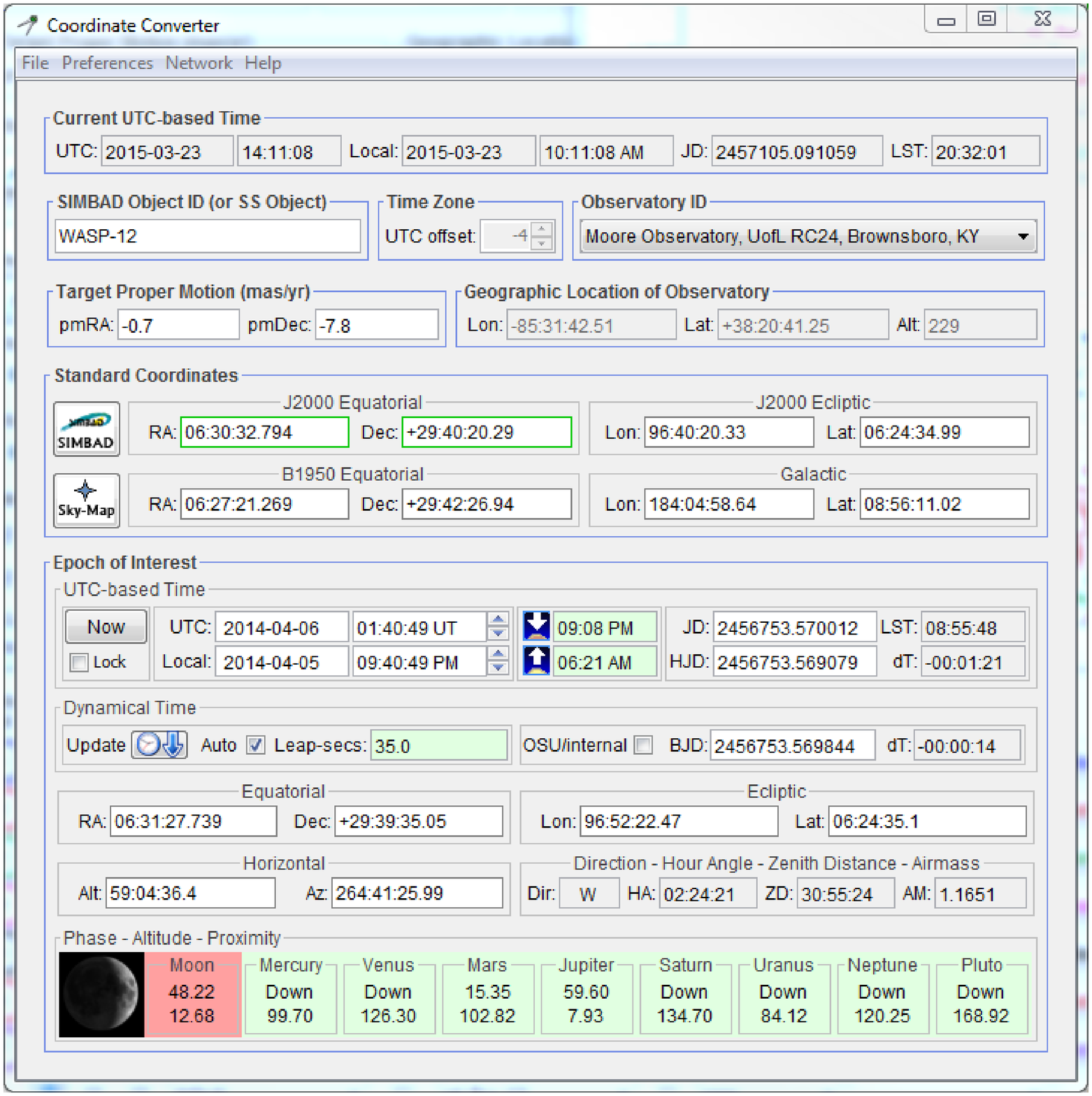} }
\caption{The \textit{Coordinate Converter} panel. The displayed state of the panel is shown after entering WASP-12 in the SIMBAD Object ID field, selecting the Observatory ID, and entering the date and time as UTC 2014-04-06 01:40:49. All other coordinate formats, time formats, solar system object proximities and altitudes, and moon phase are automatically calculated. Note that the red background of the Moon proximity box indicates that the moon is less than $15\degrees$ from the target ($12.68\degrees$ in this case). The active coordinate source used to calculate the other coordinate formats has a green border (J2000 Equatorial in this case). Any coordinate or time format can become the active source by directly typing a value into the field and pressing \textit{$<$Enter$>$}. See text for more details.}
\label{fig:aijccpanel}
\end{center}
\end{figure*}

The user controlled instance of the \textit{Coordinate Converter} panel is started by clicking the CC icon $\left(~\includegraphics[scale=0.37, trim=1.7mm 2mm 1.7mm 0mm]{aijccicon}~\right)$ on the AIJ Toolbar. Figure \ref{fig:aijccpanel} shows the panel after entering WASP-12 in the SIMBAD Object ID field, selecting the Observatory ID, and entering the date and time as UTC 2014-04-06 01:40:49. All other coordinate formats, time formats, solar system object proximities and altitudes, and moon phase are automatically calculated. Note that the red background of the Moon proximity box indicates that the moon is less than $15\degrees$ from the target ($12.68\degrees$ in this case). 

The active coordinate source used to calculate the other coordinate formats has a green border (J2000 Equatorial in this case). Any coordinate or time format can become the active source by directly entering a value into the field and pressing \textit{$<$Enter$>$}. The two fields in the \textit{UTC-based Time} sub-panel shown in Figure \ref{fig:aijccpanel} with a green background show the time of PM nautical twilight (upper) and AM nautical twilight (lower) for the date, time, and observatory location specified. If the date and time showing in the other \textit{UTC-based Time} fields is during dark time, the two twilight field backgrounds are green, otherwise they are gray.

The $\bjdtdb$ time format requires dynamical time as the time-base. Dynamical time accounts for the changing rotational speed of the Earth by implementing leap-seconds. Leap-second updates are not periodic, but are announced six months before taking effect. The U.S. Naval Observatory website posts a list of all leap seconds along with the effective date of each one. To ensure that conversion to $\bjdtdb$ time format is accurate, CC's leap second table should be updated by clicking the \textit{Update} button in the \textit{Dynamic Time} sub-panel every 6 months or so while connected to the internet. 

DP creates an instance of CC (DPCC), and depending on user settings, data can be extracted from FITS header information or entered manually to control the settings used by DPCC to calculate new astronomical values to add to the calibrated image's FITS header information. If the FITS header contains the time of observations, the target's SIMBAD ID or coordinates, and the observatory's ID or coordinates, DPCC calculations can be executed with no user input. If target and/or observatory information is not available in the header, that missing information can be entered by the user as previously shown in Figure \ref{fig:aijccpanel}. Appendix \ref{sec:fitsheaderupdates} describes how to set up the various DP FITS header update options and modes of operation.

\begin{figure}
\begin{center}
\resizebox{\columnwidth}{!}{
\includegraphics*{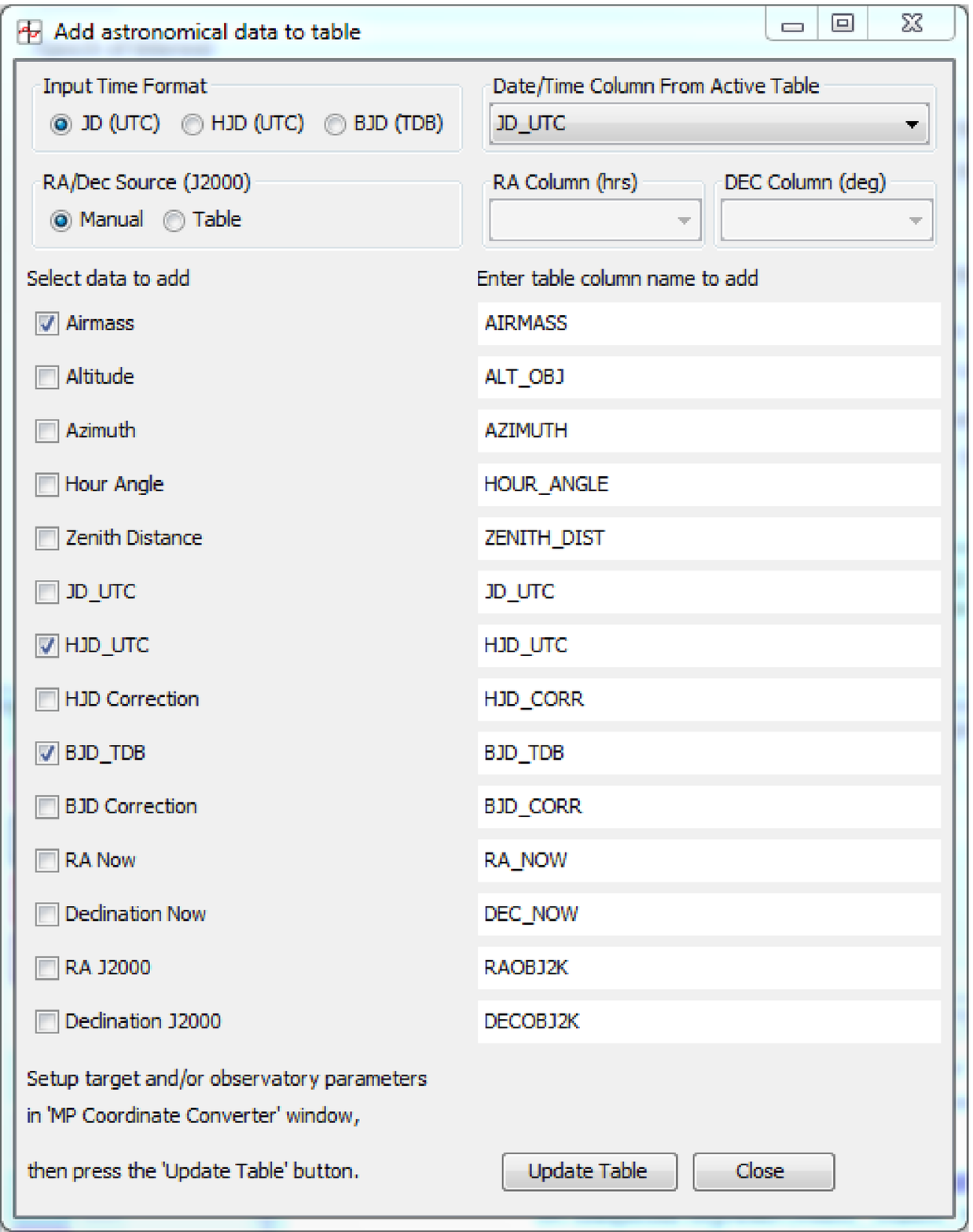} }
\caption{The \textit{Add astronomical data to table} panel. The user sets the time format and data column name to be used for extraction of time from each row in the measurements table. An instance of CC referred to as MPCC is also opened along with the add data panel. In the \textit{RA/Dec Source} \textit{Manual} mode shown here, the target coordinates and observatory location should be manually entered into the MPCC panel. Then, when the \textit{Update Table} button is clicked, the settings shown here will cause new data columns AIRMASS, HJD\_UTC, and BJD\_TDB to be calculated and added to the measurements table.}
\label{fig:aijadddatapanel}
\end{center}
\end{figure}

The menu item at \textit{Multi-plot Main}$\rightarrow$\textit{Table}$\rightarrow$\textit{Add new astronomical data columns to table} opens the panel shown in Figure \ref{fig:aijadddatapanel}. In that panel, the user sets the time format and data column name to be used for extraction of time from each row in the measurements table. An instance of CC referred to as MPCC is also opened along with the \textit{Add astronomical data to table} panel. With the \textit{RA/Dec Source} mode set to \textit{Manual} as shown in Figure \ref{fig:aijadddatapanel}, the target coordinates and observatory location should be manually entered into the MPCC panel. With the \textit{RA/Dec Source} mode set to \textit{Table}, the RA and Dec coordinates must have been extracted from the FITS header and added to the measurements table during photometry (see Appendix \ref{sec:photset}). In this case, the corresponding data column labels should be selected in the \textit{RA Column} and \textit{DEC Column} pull-down menus. Then, when the \textit{Update Table} button is clicked, the settings shown in Figure \ref{fig:aijadddatapanel} will cause new data columns AIRMASS, HJD\_UTC, and BJD\_TDB to be calculated and added to the measurements table. The new data columns can now be used for plotting, detrending, etc., and the updated measurements table can be saved to disk.

\subsection{FITS Header Editor}\label{sec:fitsheadereditor}

\begin{figure*}
\begin{center}
\resizebox{\textwidth}{!}{
\includegraphics*{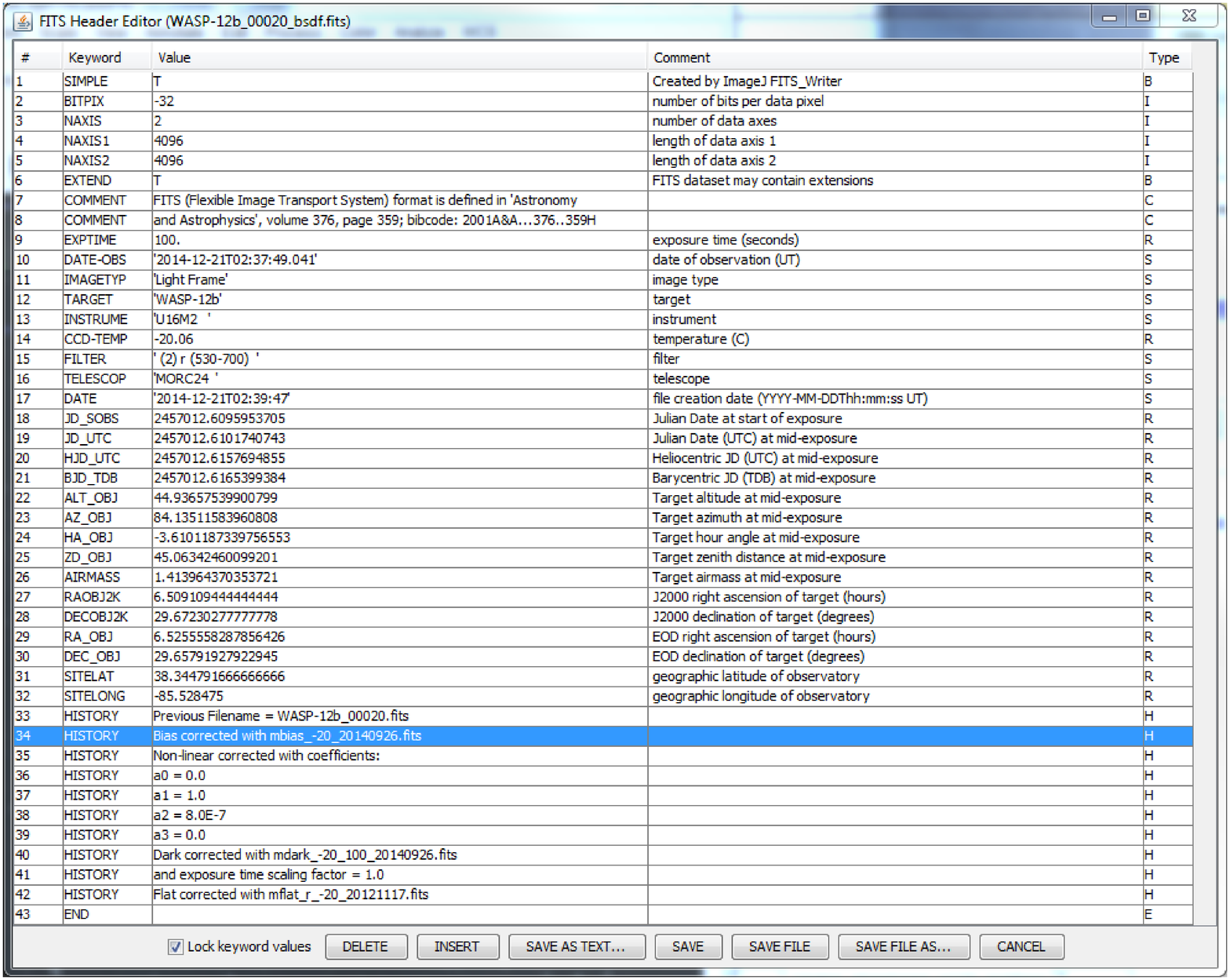} }
\caption{The \textit{FITS Header Editor} panel. Most fields can be directly edited by double-clicking in the fields and editing the contents using the keyboard. The \textit{DELETE} button at the bottom of the display will delete the selected row. The \textit{INSERT} button adds a new row under the currently selected row. The header can be saved to a text file, back to the image header in memory, or directly to disk. See the text for more details and a list of fields that cannot be directly edited.}
\label{fig:aijfitsheadereditor}
\end{center}
\end{figure*}

Information contained in the header of a FITS image open in AIJ can be displayed and optionally edited by clicking the \textit{FITS Header Editor} icon $\left(\,\includegraphics[scale=0.37, trim=1.7mm 2.2mm 1.5mm 0mm]{aijfitsheadereditoricon}\,\right)$ above an image display. The  \textit{FITS Header Editor} panel shown in Figure \ref{fig:aijfitsheadereditor} opens. A FITS header is made up of keywords and associated value and comment fields. FITS header keywords and values should not be edited unless the user understands the impact the change may have on the downstream interpretation of the image data. 

A header value is edited by double-clicking in the field and editing the value using the keyboard. The keyword values are locked by default, but can be edited after deselecting the \textit{Lock Keyword Values} checkbox. The \textit{Value} field may contain a string (i.e. text enclosed in single quotes), integer, real number, or boolean (i.e. a T or F). AIJ checks the formatting of the \textit{Value} field to ensure that the new entry meets the FITS specification for one of the data types allowed. The \textit{Type} field is automatically set based on contents of the \textit{Value} field and cannot be directly edited. Rows with keyword values SIMPLE, BITPIX, NAXIS, NAXIS1, and NAXIS2 can not be edited since these values are automatically set by AIJ according to the image's characteristics. The END keyword must always be present in the last row and cannot be edited.

The buttons along the bottom of the editor, from left to right, allow the user to delete the selected row of data (which shows as highlighted in blue) from the header, insert a new row below the currently selected row, save the contents of the FITS header to a text file, save the new header to the image in memory, save the image and new header to disk (and memory) using the same filename, save the image and new header to disk using a new filename, or cancel the changes and exit the editor.

\subsection{Astrometry/Plate Solving}\label{sec:astrometry}

The astrometry feature ``plate solves'' images using an internet connection to the astrometry.net web portal at \url{nova.astrometry.net} \citep{Lang:2010}. After a successful astrometric solution is found, WCS headers are automatically added to the FITS image header, and the file can optionally be resaved with the new headers. AIJ searches the image and extracts the source locations. Only the $x,y$ coordinates for a subset of the brightest extracted sources are sent to \url{nova.astrometry.net}. The actual image is not transferred across the network, which limits network traffic and improves the solve time.

\begin{figure*}
\begin{center}
\resizebox{\textwidth}{!}{
\includegraphics*{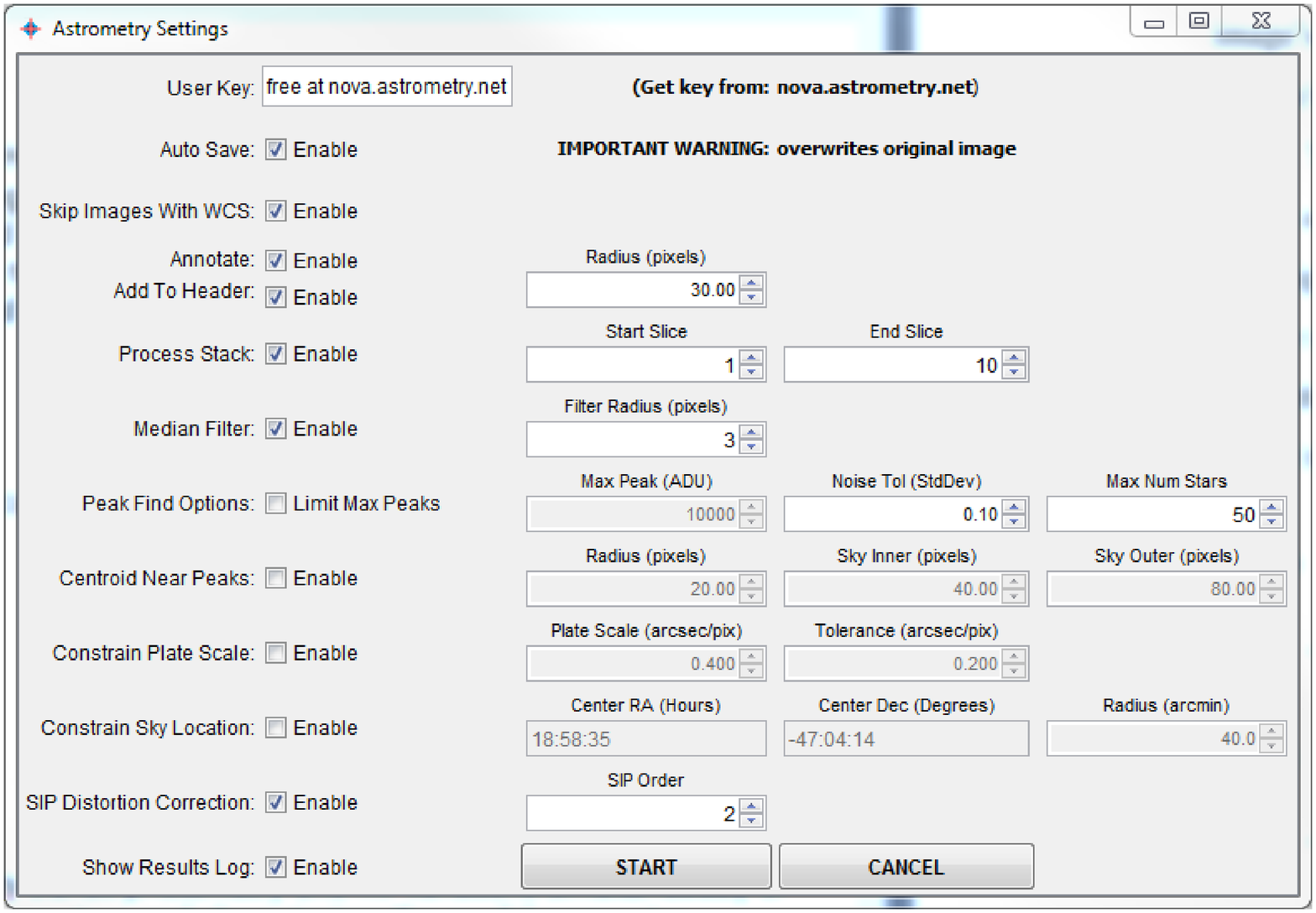} }
\caption{The \textit{Astrometry Settings} panel. The astrometry feature ``plate solves'' images using an internet connection to the astrometry.net web portal at \url{nova.astrometry.net}. After a successful astrometric solution is found, WCS headers will automatically be added to the FITS image header and the file can optionally be automatically resaved with the new headers. See text for more details.}
\label{fig:aijastrometrypanel}
\end{center}
\end{figure*}

A left-click on the \textit{Astrometry} icon $\left(\,\includegraphics[scale=0.37, trim=1.7mm 2.2mm 1.5mm 0mm]{aijastrometryicon}\,\right)$ above an image opens the \textit{Astrometry Settings} panel shown in Figure 
\ref{fig:aijastrometrypanel}. A right-click on the \textit{Astrometry} icon starts the plate solve process using the previous settings panel values. A left-click on the \textit{Astrometry} icon, after the plate solve process has started, aborts the process. DP also provides an option in the \textit{FITS Header Updates} sub-panel (see Figure \ref{fig:aijdataprocessor}) to plate solve each image as part of the calibration process. A free user key must be obtained from \url{nova.astrometry.net} and entered into the \textit{User Key} field of the \textit{Astrometry Settings} panel to enable the astrometry feature.

Images can be blindly solved with no knowledge of the sky coordinates or plate scale of the image. The default settings shown in Figure \ref{fig:aijastrometrypanel} should work for most images. Solve time may be faster if the \textit{Plate Scale} is known and entered into the field on the set up panel. If the approximate sky coordinates of the center of the image are known, entering the \textit{Center RA} and \textit{Center Dec} values may also improve solve time. However, the search \textit{Radius} must be at least as large as the field of view in the image. If an image has been defocused to improve photometric precision, the \textit{Centroid Near Peaks} option may improve the determination of the location of the center of each source.

When the \textit{START} button is clicked, the set up panel closes and the plate solve process starts. Progress is shown in the lower half of the AIJ Toolbar. By default, a log file is created to record the results of each plate solve. If a time-series of images have been opened into an AIJ image stack, the full set of images can be solved by selecting the \textit{Process Stack} option. The entire plate solve process takes $\sim10-20$ seconds per image.

When a field is successfully solved, \url{nova.astrometry.net} returns a list of sources that are in the image. The source names can be displayed in the image by enabling \textit{Annotate} and/or saved to the FITS header by enabling \textit{Add to Header}. More set up information is provided in the tool-tip help that is available for each option in the user interface panel. 

\subsection{Image Alignment}\label{sec:stackaligner}

\begin{figure}
\begin{center}
\resizebox{\columnwidth}{!}{
\includegraphics*{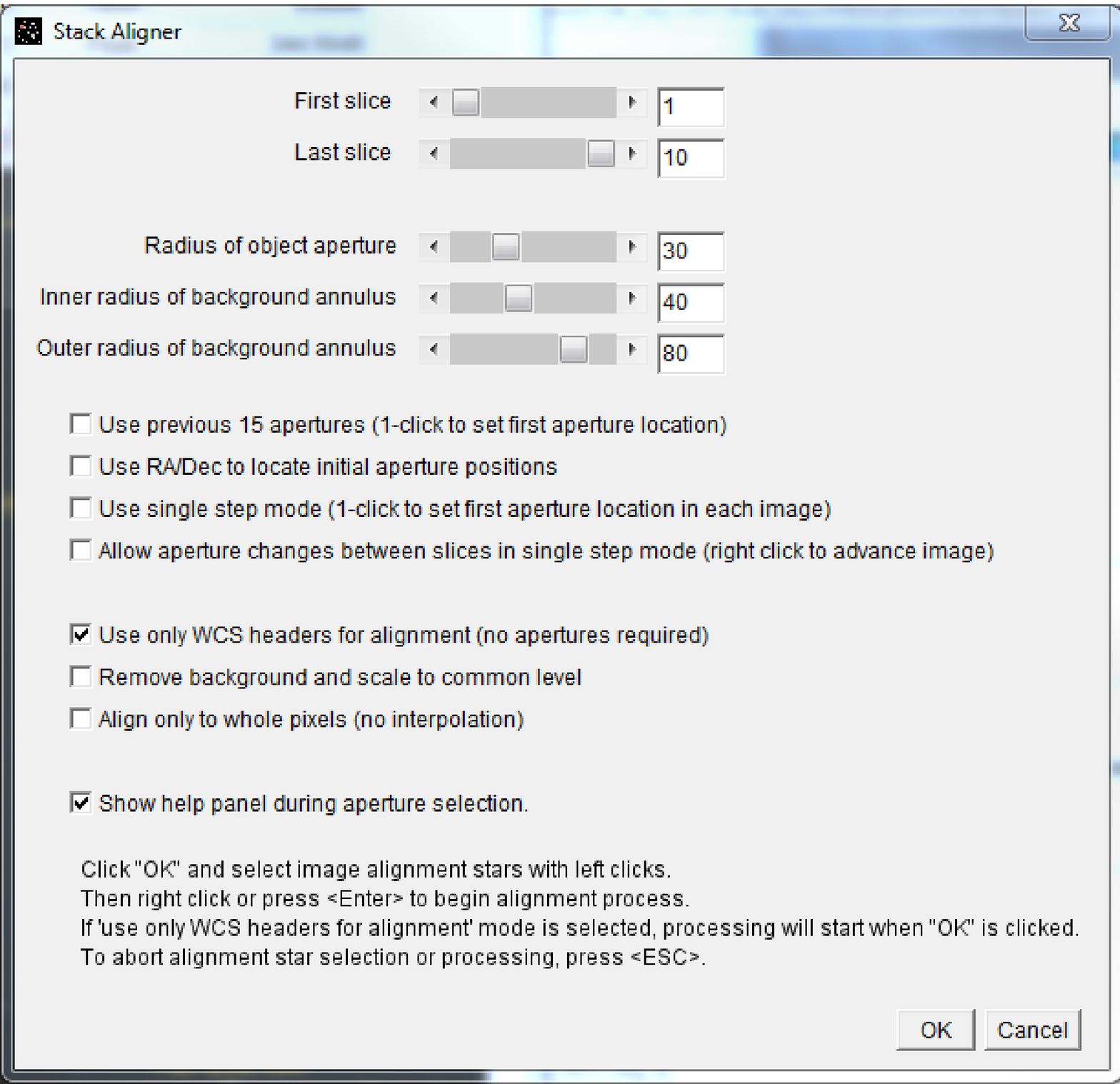} }
\caption{The \textit{Stack Aligner} panel. Images can be aligned using information in the WCS headers, if images have been plate solved. Otherwise, apertures are placed around selected alignment stars, and images are aligned based on the centroid offsets between consecutive images. See text for more details.}
\label{fig:aijstackaligner}
\end{center}
\end{figure}

The images within a stack can be aligned using the \textit{Stack Aligner} icon $\left(\,\includegraphics[scale=0.38, trim=1.7mm 1.8mm 1.5mm 0mm]{aijstackalignericon}\,\right)$ above an image stack. The \textit{Stack Aligner} panel shown in Figure \ref{fig:aijstackaligner} opens and provides two methods to align images. At the time of writing, Stack Aligner only supports image translation for alignment. Image rotation and scaling are not currently implemented. 

If all images in the stack have been plate solved, the images can be aligned using information in the WCS headers. To use this mode, enable the \textit{Use only WCS headers...} option and click the OK button to start the alignment process. All images in the stack will then be aligned to the first image.

If images have not been plate solved, apertures may be used to identify alignment stars. Aperture placement is performed in the same way as described for Multi-Aperture in \S \ref{sec:multiaperture}, and images are aligned based on the average of alignment star centroid offsets between consecutive images. Aperture alignment works best when at least a few ($\sim3-5$) isolated stars are available in the images. Aperture alignment will fail if the shift from one image to the next is larger than the aperture radius. However, the aperture radius can be made arbitrarily large as long as centroid doesn't capture a neighboring star instead of the alignment star. For cases with large image shifts, the \textit{Use single step mode} option allows the user to click on the first alignment star in each image of the sequence, avoiding apertures centroiding on the wrong star. However, this mode requires the user to click in each image of the sequence, which may become impractical for very long image sequences.

An option is available in the \textit{Process} menu above an image display that provides \textit{Image Stabilizer} functionality that is useful for processing images of non-stellar objects.  This tool will remove atmospheric jitter from a rapid sequence of planetary or lucky images, or track a comet over a long duration as it moves across a star field.

\subsection{Radial Profile}\label{sec:radialprofile}

\begin{figure}
\begin{center}
\resizebox{\columnwidth}{!}{
\includegraphics*{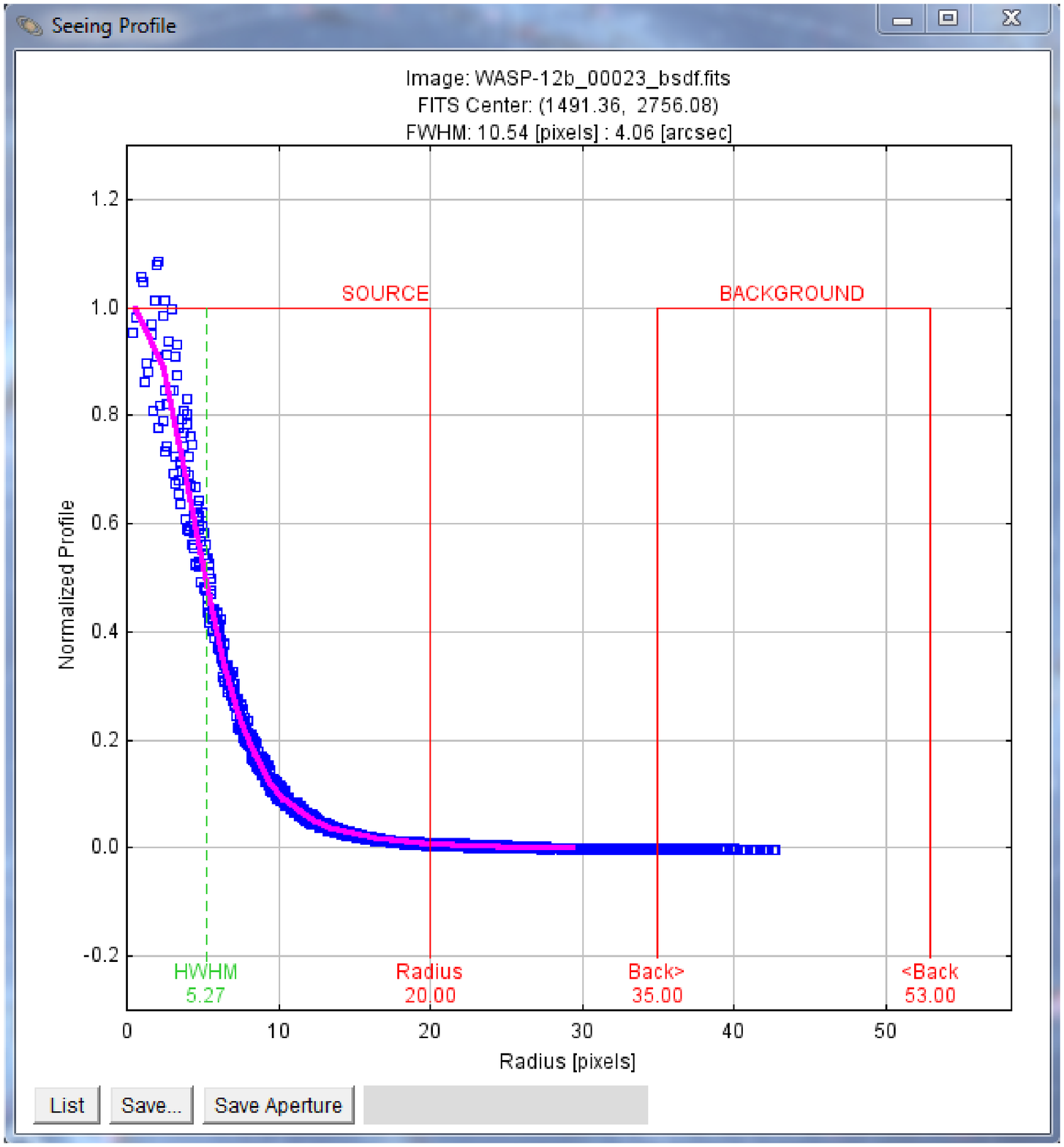} }
\caption{Radial Profile (Seeing Profile) plot. Alt-left-click near an object in an image to produce an azimuthally averaged radial profile plot. The plot shows the half-width at half-maximum (HWHM), the FWHM, and suggested aperture radii in pixels. The aperture radius is set to $1.7\times$FWHM, the inner radius of the sky-background annulus is set to $1.9\times$FWHM, and the outer radius of the sky-background annulus is set to $2.55\times$FWHM. The FWHM is also given in seconds of arc if valid WCS headers are available. The \textit{Save Aperture} button transfers the suggested aperture radii to the \textit{Aperture Photometry Settings} discussed in Appendix \ref{sec:photset}.}
\label{fig:aijseeingprofile}
\end{center}
\end{figure}

An azimuthally averaged radial profile of an object can be plotted by left-clicking near an object in an image, and then selecting \textit{Analyze}$\rightarrow$\textit{Plot seeing profile} in the menus above the image display. As a short cut, simply alt-left-click near an object in an image to produce the plot. If centroid is enabled, the radial profile will be centered on the object. Figure \ref{fig:aijseeingprofile} shows an example radial profile plot. The plot shows the half-width at half-maximum (HWHM), the FWHM, and suggested aperture radii in pixels. The aperture radius is set to $1.7\times$FWHM, the inner radius of the sky-background annulus is set to $1.9\times$FWHM, and the outer radius of the sky-background annulus is set to $2.55\times$FWHM. These radii give an equal number of pixels in the aperture and sky-background annulus. The FWHM is also given in seconds of arc if valid WCS headers are available. The \textit{Save Aperture} button transfers the suggested aperture radii to the \textit{Aperture Photometry Settings} discussed in Appendix \ref{sec:photset}. 

\subsection{Photometry Settings}\label{sec:photset}

\begin{figure*}
\begin{center}
\resizebox{\textwidth}{!}{
\includegraphics*{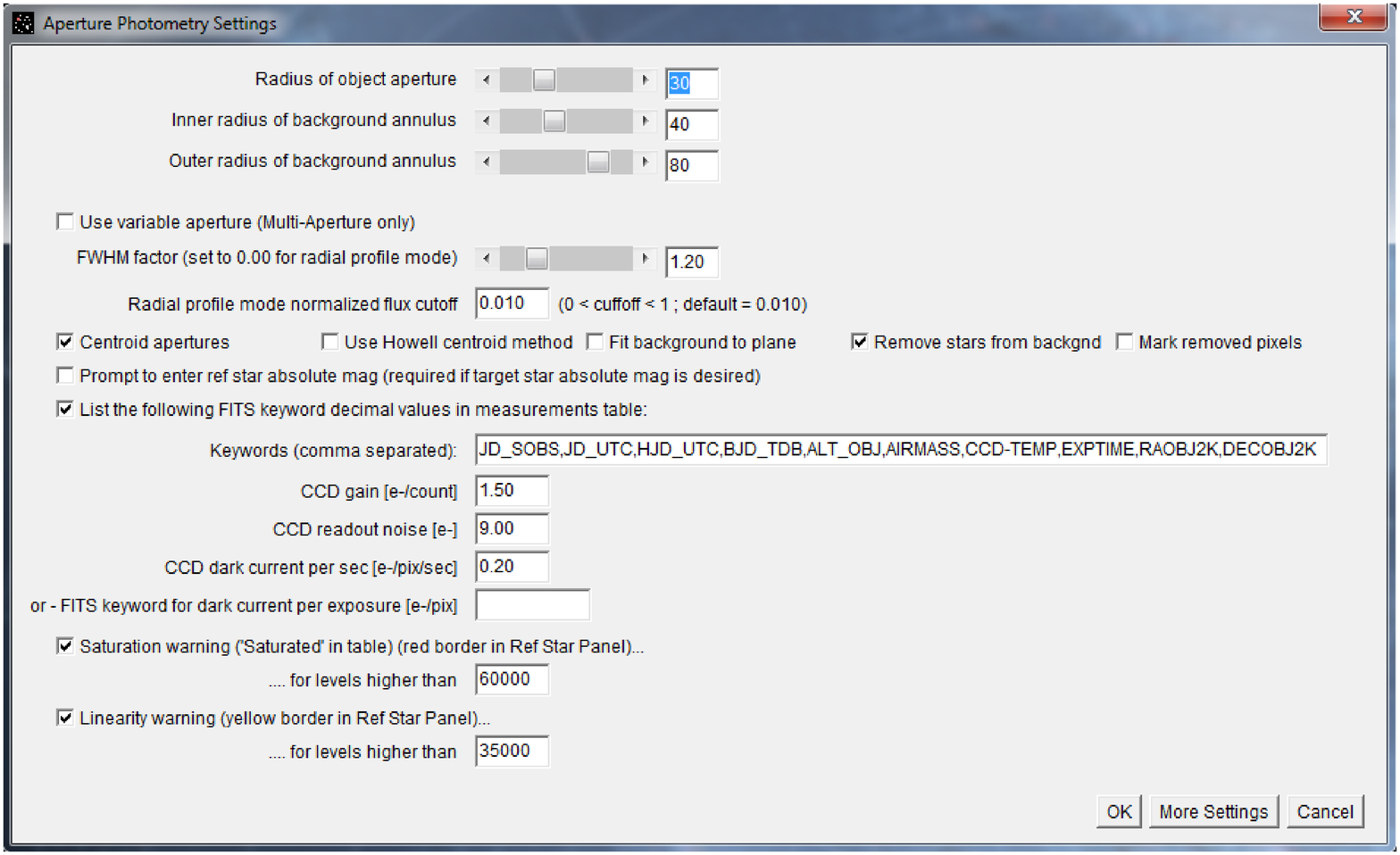} }
\caption{The \textit{Aperture Photometry Settings} panel. The panel provides access to the aperture radii, centroid, and background settings that are also available in the \textit{Multi-Aperture Measurements} set up panel. A list of FITS keywords can be entered to specify that the corresponding numeric data in the image headers be added to the measurements table. The CCD gain, readout noise, and dark current should be entered for proper photometric error calculations. Linearity and saturation warning levels should be entered so that indicators of those conditions can we properly noted in the measurements table (see Appendix \ref{app:measurementstable}) and the \textit{Multi-plot Reference Star Settings} panel (see \S \ref{sec:ensemblemanagement}). The \textit{More Settings} button provides access to the additional options shown in Figure \ref{fig:aijapsetmorepanel}.}
\label{fig:aijapsetpanel}
\end{center}
\end{figure*}

\begin{figure*}
\begin{center}
\resizebox{\textwidth}{!}{
\includegraphics*{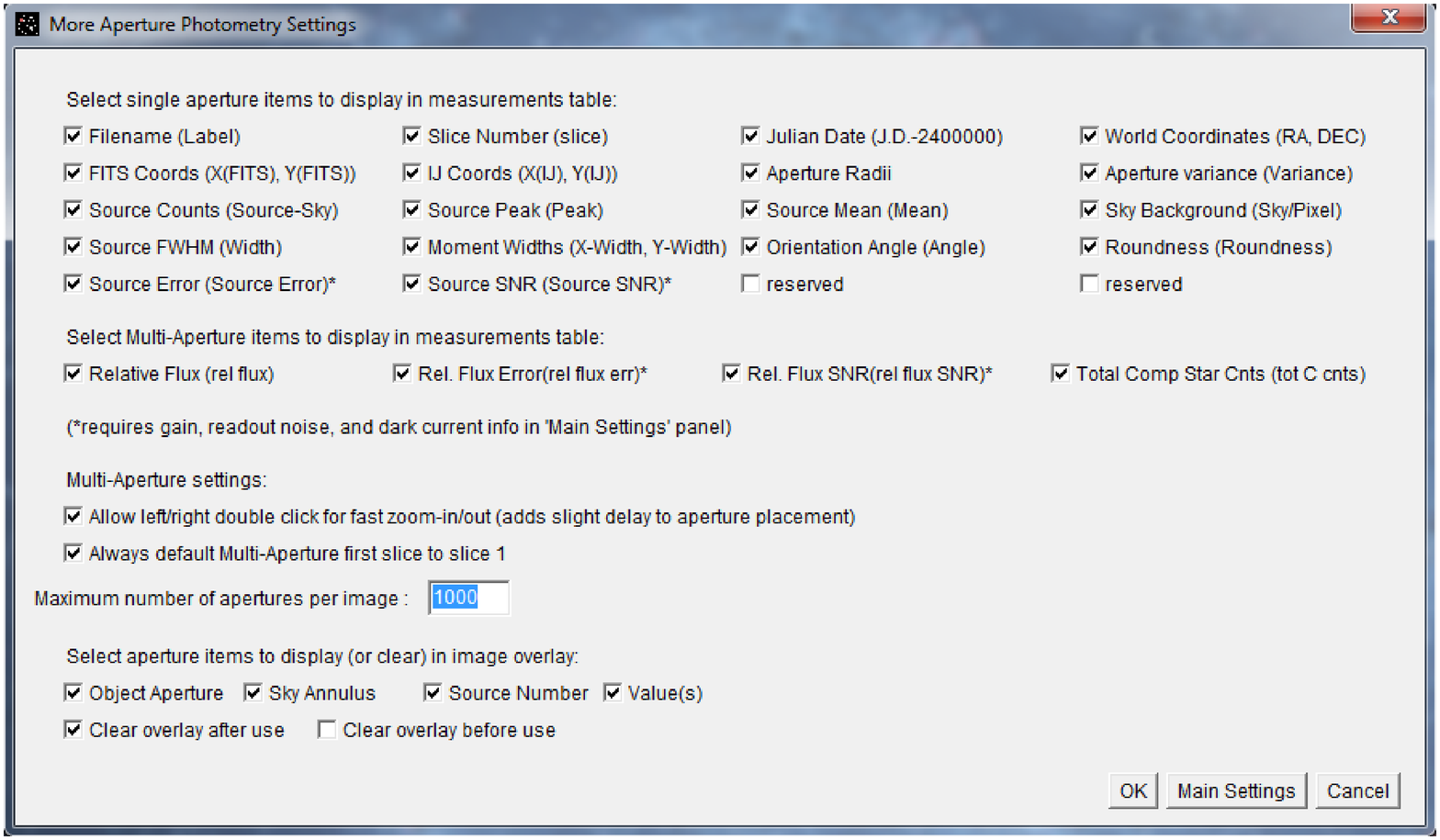} }
\caption{The \textit{More Aperture Photometry Settings} panel. The top two sections allow the selection of photometric data items to be included in the measurements table. It is highly recommended to enable all data items since some AIJ functionality requires certain data items to be available in the table. The maximum number of apertures per image is set high by default, but can be changed as needed.}
\label{fig:aijapsetmorepanel}
\end{center}
\end{figure*}

Settings related to photometric measurements are specified in the \textit{Aperture Photometry Settings} panel shown in Figure \ref{fig:aijapsetpanel} and the \textit{More Aperture Photometry Settings} panel shown in Figure \ref{fig:aijapsetmorepanel}. The \textit{Aperture Photometry Settings} panel can be accessed by double-clicking the \textit{Single Aperture Photometry} icon $\left(\,\includegraphics[scale=0.37, trim=1.7mm 2.2mm 1.3mm 0mm]{aijapmodeicon}\,\right)$ on the AIJ Toolbar, by clicking the set aperture icon $\left(\,\includegraphics[scale=0.37, trim=1.5mm 2.6mm 1.5mm 0mm]{aijapseticon}\,\right)$ above an image, by clicking the \textit{Aperture Settings} button near the bottom of the \textit{Multi-Aperture Measurements} set up panel, or by clicking the set aperture icon $\left(\,\includegraphics[scale=0.35, trim=1.4mm 2.6mm 1.5mm 0mm]{aijdpapseticon}\,\right)$ in the DP \textit{Control Panel} sub-panel. The \textit{More Aperture Photometry Settings} panel is accessed by clicking the \textit{More Settings} button on the \textit{Aperture Photometry Settings} panel.

The \textit{Aperture Photometry Settings} panel provides access to the aperture radii, centroid, and background settings that are also available in the \textit{Multi-Aperture Measurements} set up panel (see \S \ref{sec:multiaperture}). A list of FITS keywords can be entered to specify that the corresponding numeric data in the image headers be added to the measurements table. The CCD gain, readout noise, and dark current should be entered for proper photometric error calculations. Linearity and saturation warning levels should be entered so that indicators of those conditions can we properly noted in the measurements table (see Appendix \ref{app:measurementstable}) and the \textit{Multi-plot Reference Star Settings} panel (see \S \ref{sec:ensemblemanagement}). 

The \textit{Howell} centroid method \citep{Howell:2006} gives highly repeatable \textit{x,y} results (i.e. is not sensitive to the \textit{x,y} starting location). If the \textit{Howell} option is disabled, a center-of-mass (i.e. center-of-flux) routine is used to calculate the centroid. This method provides better results when placing apertures around defocused stars.   

The top two sections of the \textit{More Aperture Photometry Settings} panel allow the selection of photometric data items to be included in the measurements table. See Appendix \ref{app:measurementstable} for a description of each data item. It is highly recommended to enable all data items since some AIJ functionality requires certain data items to be available in the table. The maximum number of apertures per image is set high by default, but can be changed as needed. The aperture display options control which parts of the aperture overlay are displayed on the image. The \textit{Sky Annulus}, \textit{Source Number}, and \textit{Value} check boxes duplicate the same controls available above an image display. For typical operation, \textit{Clear overlay after use} should be selected and \textit{Clear overlay before use} should be deselected.

\subsection{Data Processor FITS Header Updates}\label{sec:fitsheaderupdates}

\begin{figure}
\begin{center}
\resizebox{\columnwidth}{!}{
\includegraphics*{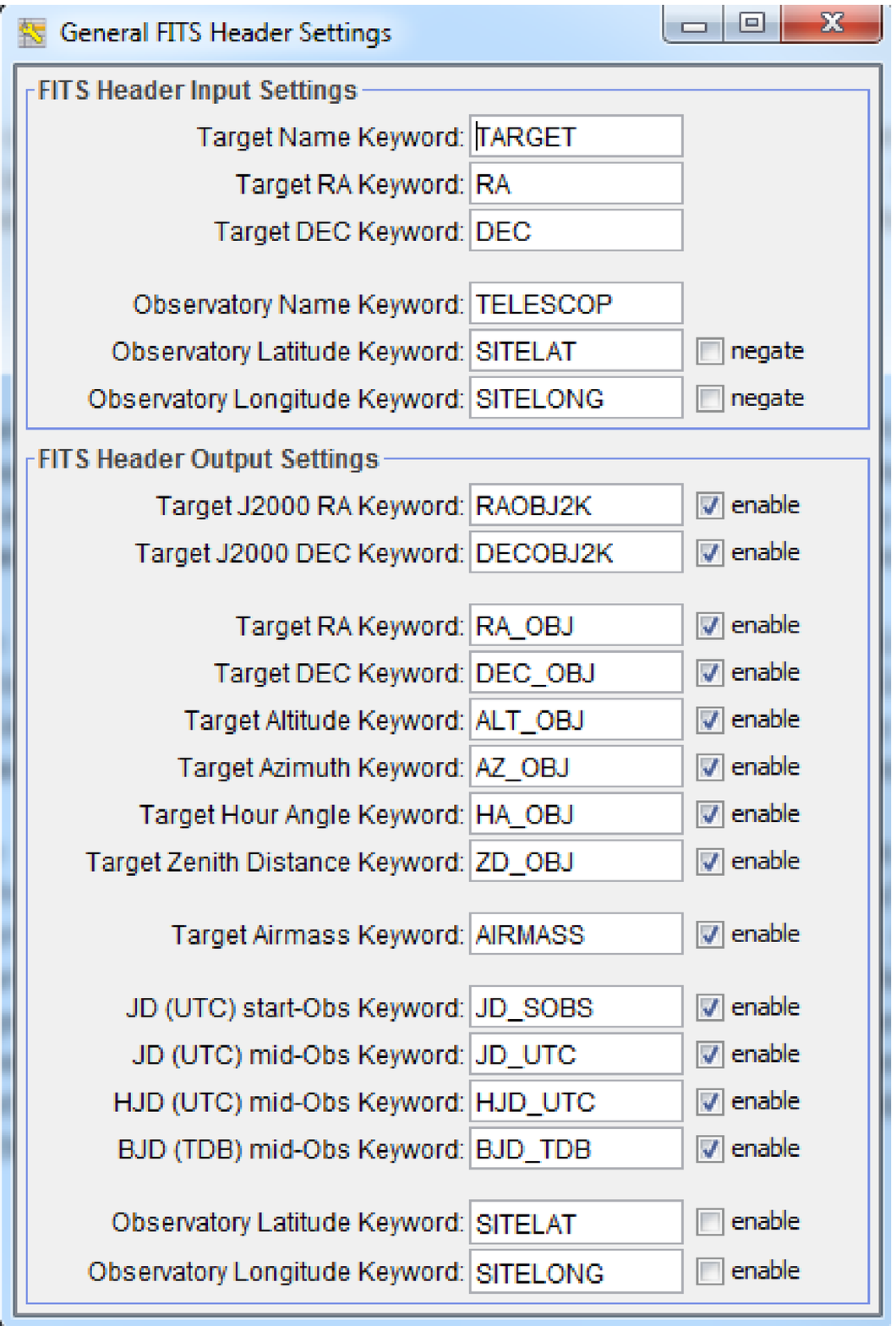} }
\caption{The \textit{General FITS Header Settings} panel. The keywords specified in the \textit{FITS Header Input Settings} sub-panel can be used to automatically supply the target and observatory coordinates to DPCC. The \textit{FITS Header Output Settings} sub-panel specifies new astronomical values to calculate and keyword names to use to store those values in the FITS header of the calibrated image.} 
\label{fig:aijfitsheaderupdates}
\end{center}
\end{figure}

DP provides the option to calculate new astronomical data and add it to the FITS header of a calibrated image. The option is enabled by selecting the \textit{General} check box in the \textit{FITS Header Updates} sub-panel of DP. A set up panel is accessed by clicking the \textit{General FITS Header Settings} icon $\left(\,\includegraphics[scale=0.35, trim=1.5mm 2.7mm 1.5mm 0mm]{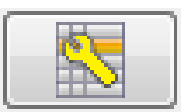}\,\right)$. The \textit{General FITS Header Settings} panel shown in Figure \ref{fig:aijfitsheaderupdates} opens. The \textit{FITS Header Input Settings} sub-panel defines FITS header keywords that are preexisting in the raw science image headers that may be used in the calculation of various new astronomical data values specified in the \textit{FITS Header Output Settings} sub-panel. These new astronomical data are added to the FITS header of the calibrated images using the specified keyword names.

The astronomical calculations require the target sky coordinates and observatory geographical coordinates to be supplied to DP Coordinate Converter (DPCC; see Appendix \ref{sec:coordinateconverter}). Both sets of coordinates can be manually entered into DPCC by setting the \textit{Target Coordinate Source} option to \textit{Coordinate Converter manual entry} and the \textit{Observatory Location Source} option to \textit{Coordinate Converter manual entry}. Alternatively, coordinate information can be automatically supplied to DPCC from the FITS header of the raw image by setting the corresponding keyword names in the \textit{FITS header Input Settings} sub-panel and then selecting the appropriate \textit{Target Coordinate Source} and \textit{Observatory Location Source} to match the type of data in the header (i.e. a target or observatory name or a set of numeric coordinates). 

\subsection{Save All}\label{sec:saveall}

\begin{figure}
\begin{center}
\resizebox{\columnwidth}{!}{
\includegraphics*{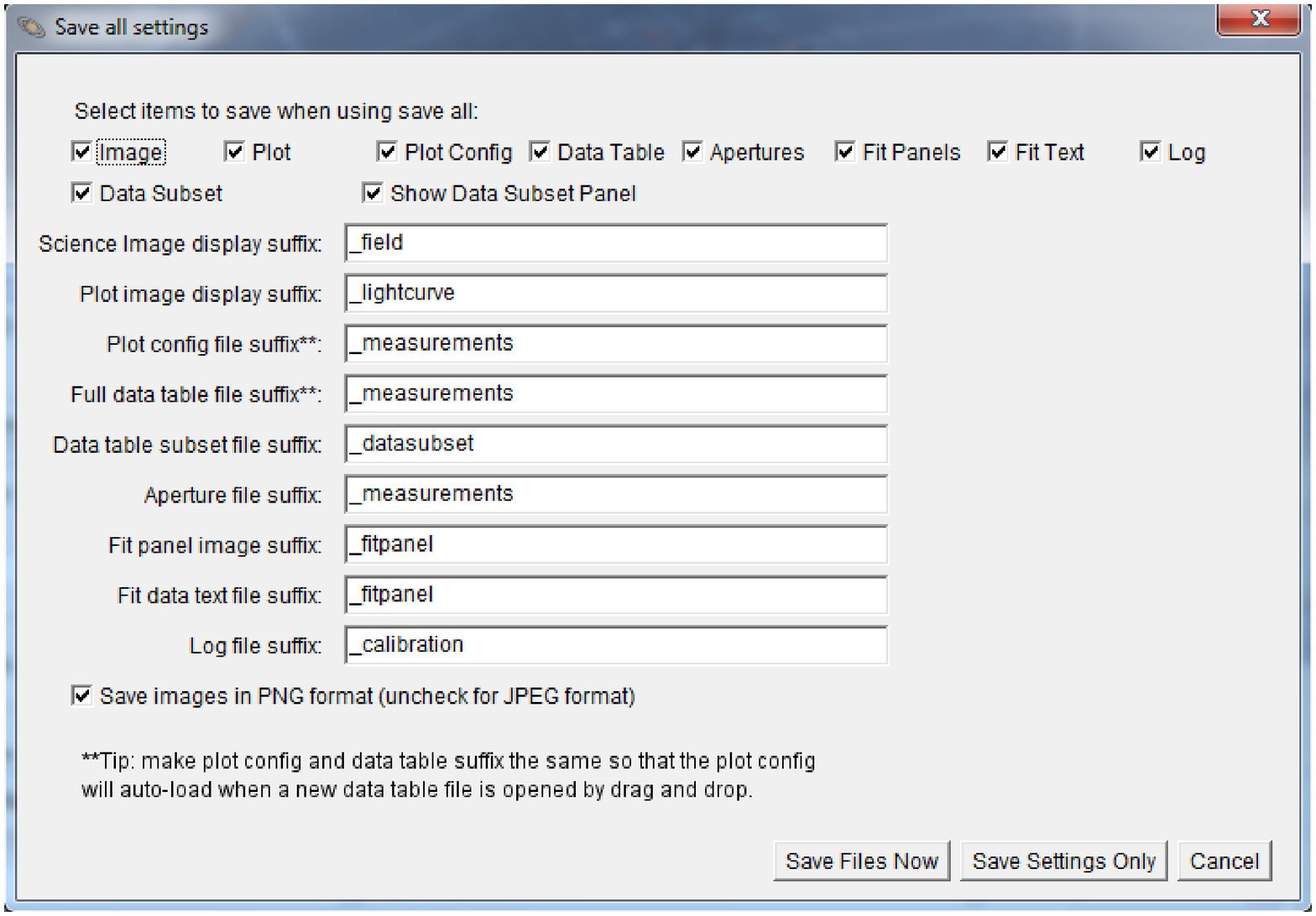} }
\caption{The \textit{Save All} panel. All typical data and image products produced by AIJ can be saved with one action using the \textit{Save All} feature. Individual items that can be included in the save all process are enabled in the array of checkboxes at the top of the panel. A base file path is specified in a separate file save dialog and then the suffix names specified in the \textit{Save All} panel are appended as the files are saved. The checkbox at the bottom of the panel specifies whether images are saved in \textit{png} or \textit{jpeg} format.} 
\label{fig:aijsaveallpanel}
\end{center}
\end{figure}

All typical data and image products created by AIJ can be saved with one action using the \textit{Save All} features in the \textit{File} menu of the \textit{Multi-plot Main} panel. There are also \textit{File} menu options in \textit{Multi-plot Main} and image display panels to save each data product separately. The menu option at \textit{File}$\rightarrow$\textit{Save all (with options)} opens the \textit{Save All} settings panel shown in Figure \ref{fig:aijsaveallpanel}.  Individual items that can be included in the save all process are enabled in the array of checkboxes at the top of the panel. A base file name, \textit{filebase}, is specified in a separate \textit{file save} dialog and then the suffix names specified in the \textit{Save All} panel are appended as the files are saved. The checkbox at the bottom of the panel specifies whether images are saved in \textit{png} or \textit{jpg} format. The menu option at \textit{File}$\rightarrow$\textit{Save all} runs the process using previous settings without opening the \textit{Save All} settings panel.

If the \textit{Image} option is enabled, the image in the active image display (i.e. the last image clicked with the mouse if more than one is open) is saved to a file name constructed from the \textit{filebase} and the suffix specified in \textit{Science image display suffix} and with a filetype of \textit{png} or \textit{jpg}, depending on the format selected at the bottom of the panel. For the configuration shown in Figure \ref{fig:aijsaveallpanel}, the file would be named ``\textit{filebase}\_field.png''. This image will include any items currently being displayed in the non-destructive image overlay. If no image display is open, no file will be created. If the \textit{Plot} option is enabled and a plot is being displayed, the plot image will be saved using the \textit{filebase} and file suffix in \textit{Plot image display suffix} and with a filetype of either \textit{png} or \textit{jpg}.

If the \textit{Plot Config} option is enabled, the current state of all plot settings will be saved to a plot configuration file using the \textit{filebase} and file suffix in \textit{Plot config file suffix} and with a filetype of \textit{plotcfg}. If \textit{Data Table} is enabled, the full measurements table will be saved to a file using the \textit{filebase} and file suffix in \textit{Full data table file suffix} and with a filetype as specified in the AIJ Toolbar menu item \textit{Edit}$\rightarrow$\textit{Options}$\rightarrow$\textit{Input/Output}$\rightarrow$\textit{File extension for tables}. The default filetype is \textit{xls}, but this can be changed to something more unique to AIJ (e.g \textit{tbl}) so that measurements tables can be set to open into AIJ with a double-click on the corresponding file in an OS window. \textit{Tip:} If \textit{Plot config file suffix} and \textit{Full data table file suffix} are the same, when a measurements table is opened into AIJ by dragging it from the OS onto a Multi-Plot panel, or by double-clicking the file in an OS window, the plot configuration file will be loaded along with the measurements table so that the corresponding plot will be automatically recreated.

\begin{figure}
\begin{center}
\resizebox{\columnwidth}{!}{
\includegraphics*{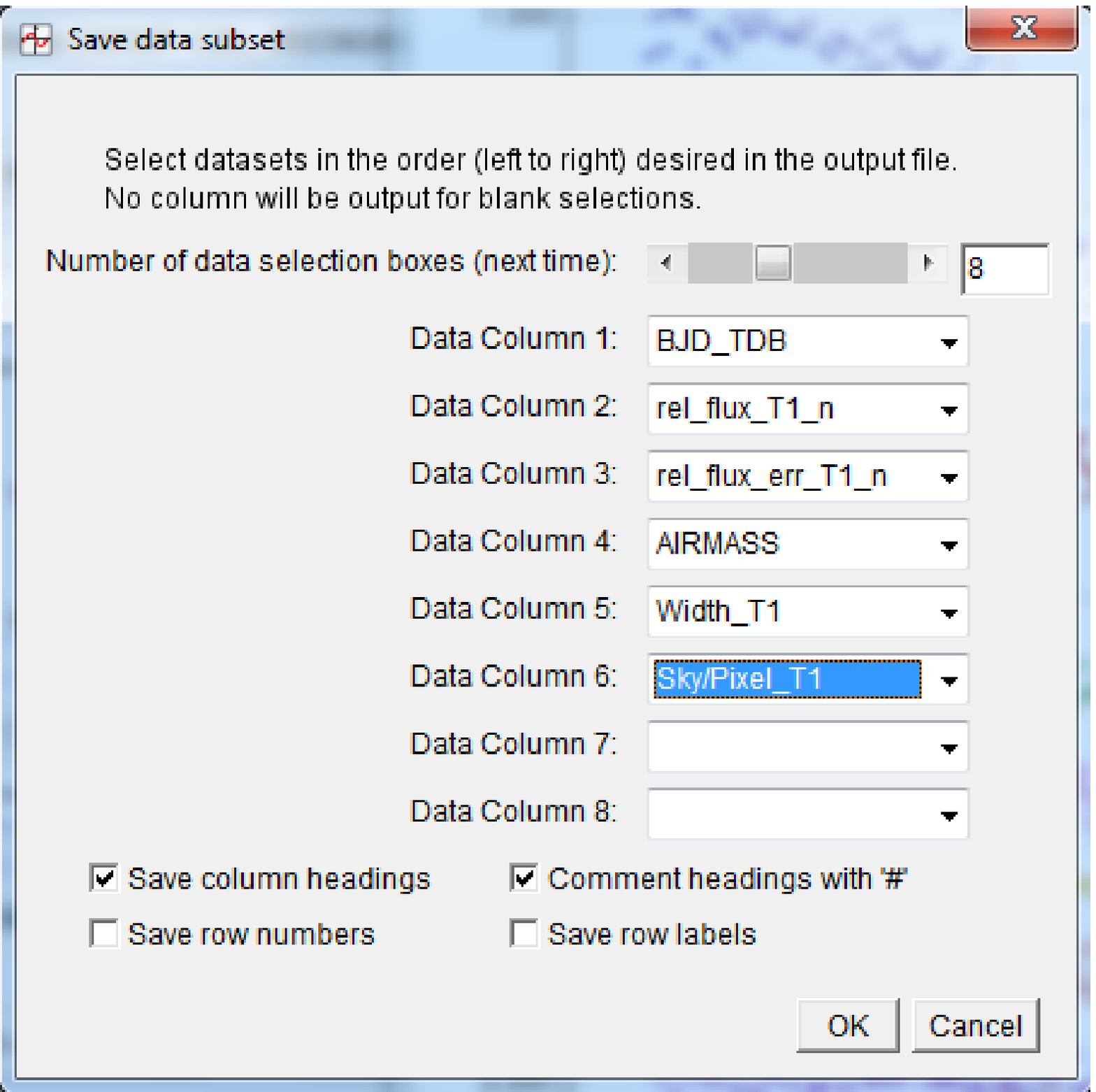} }
\caption{The \textit{Save Data Subset} panel. A subset of data columns from the measurements table can be specified and saved as a tab-delimited file. The labels of data columns that are to be included are selected in the pull-down menus. The number of available data column selections to be shown the next time the panel is opened can be changed. The options at the bottom of the panel specify if column headings, column headings hash-tag, row numbers, and/or row labels are included in the data subset file.} 
\label{fig:aijsavedatasubsetpanel}
\end{center}
\end{figure}

If the \textit{Show Data Subset Panel} and \textit{Data Subset} options are enabled, the panel shown in Figure \ref{fig:aijsavedatasubsetpanel} opens as part of the \textit{Save All} process to allow the user to select a subset of data columns from the measurements table to be saved as a separate tab-delimited file. The labels of data columns that are to be included are selected in the pull-down menus. The top-most item appears as the first column in the saved file. The bottom-most item appears as the last column in the table. Blank entries do not produce a data column. The table will be saved using the \textit{filebase} and file suffix in \textit{Data table subset file suffix} and with a filetype of \textit{dat}. If the \textit{Data Subset} option is enabled, but the \textit{Show Data Subset Panel} option is disabled, a data subset file will be created using the previous settings without opening the \textit{Save Data Subset} panel.

If the \textit{Apertures} option is enabled, the aperture settings used in the most recent Multi-Aperture run are saved to a file using the \textit{filebase} and suffix in \textit{Aperture file suffix} and with a filetype of \textit{apertures}. If the \textit{Fit Panels} option is enabled, images of all open \textit{Fit Settings} panels are saved to files using the \textit{filebase}, the suffix in \textit{Fit panel image suffix}, the \textit{Data Set} number from the \textit{Multi-Plot Y-data} panel, plus the fitted data column name, and with a filetype of either \textit{png} or \textit{jpg}. If the \textit{Fit Text} option is enabled, text-based representations of all open \textit{Fit Settings} panels are saved to files using the \textit{filebase}, the suffix in \textit{Fit data text file suffix}, the \textit{Data Set} number from the \textit{Multi-Plot Y-data} panel, plus the fitted data column name, and with a filetype of \textit{txt}. If the \textit{Log} option is enabled and a \textit{log} panel is open, the log text is saved to a file using the \textit{filebase} and suffix in \textit{Log file suffix} and with a filetype of \textit{log}.

\section{Photometric Error Calculation}\label{app:photerr}

Proper estimation of the uncertainty (i.e. error or noise) in each photometric measurement is important for reporting the significance of the measurement and plotting error bars in the light curve plot, but it is also important for the proper calculation of the best fit model to the data, since the uncertainty of each measurement, $\sigma$,  is part of the $\chisq$ calculation used in the fitting process (e.g. see equation \ref{eq:chi2fordetrend}). In short, the $\chisq$ contribution from each data point is weighted by a factor of $1/\sigma^2$, which places more weight on data with small errors, and less weight on data with large errors.

\citet{Mortara:1981} and \citet{Howell:1989} discuss the noise contributions to the measurement of a point source using CCD aperture photometry and develop the ``CCD equation'' to estimate the signal-to-noise ratio of a measurement. \citet{Merline:1995} construct a computer model of the same measurement and develop the more rigorous ``revised CCD equation''. The equation gives the total noise N in ADU for a CCD aperture photometry measurement as:
\begin{equation}
N=\frac{\sqrt{GF_*+n_{pix}(1+\frac{n_{pix}}{n_b})(GF_S+F_D+F^2_R+G^2\sigma^2_f)}}{G},
\label{eq:ccdnoise}
\end{equation}
\noindent where $G$ is the gain of the CCD in electrons/ADU, $F_*$ is the net (background subtracted) integrated counts in the aperture in ADU, $n_{pix}$ is the number of pixels in the aperture, $n_b$ is the number of pixels in the region used to estimate sky background, $F_S$ is the number of sky background counts per pixel in ADU, $F_D$ is the total dark counts per pixel in electrons, $F_R$ is read noise in electrons/pixel/read, and $\sigma_f$ is the standard deviation of the fractional count lost to digitization in a single pixel ($\sigma_f\simeq0.289$ ADU for $f$ uniformly distributed between $-0.5$ and $0.5$).

If the net integrated counts from the source, $F_*$, dominates the other terms, and the gain $G=1$, the total noise approaches the Poisson noise limit of $\sqrt{F_*}$. The noise increases as the number of pixels in the aperture increases (due to all of the secondary terms in the equation), but for a non-ideal CCD, noise introduced by inter-pixel variations decreases with increasing aperture size (except in the case of perfectly guided telescope), so the aperture size needs to be selected after considering both factors. The number of background region pixels should be chosen to be as large as possible, but not so large that the pixels far from the aperture are no longer representative of the local background at the aperture. If a source is faint relative to the local sky background, the Poisson noise of the sky background may dominate the overall measurement noise.

The AIJ photometer automatically performs the noise calculation described by equation \ref{eq:ccdnoise} for each aperture. For proper noise calculation, the gain, dark current, and read out noise of the CCD detector used to collect the data must be entered in the \textit{Aperture Photometry Settings} panel (see Appendix \ref{sec:photset}). For differential photometry, AIJ propagates the noise from all apertures to derive the error in differential flux measurements. First, the noise from the apertures of each comparison star are combined in quadrature to give the total comparison ensemble noise:
\begin{equation}
N_E=\sqrt{\sum_{i=1}^{n}N_{C_i}^2},
\label{eq:ensemblenoise}
\end{equation}
\noindent where $i$ indexes the comparison stars of the ensemble, and $N_{C_i}$ is the noise for each comparison star as calculated by equation \ref{eq:ccdnoise}, and $n$ is the number of comparison stars. Error is then propagated through the relative flux quotient to find the relative flux error for the target star as:
\begin{equation}
\sigma_{\rm{rel\_ flux}}=\frac{F_T}{F_E}\sqrt{\frac{N_T^2}{F_T^2}+\frac{N_E^2}{F_{E}^2}},
\label{eq:relfluxerr}
\end{equation}
\noindent where $F_T$ is the net integrated counts in the target aperture, $F_E$ is the sum of the net integrated counts in the ensemble of comparison star apertures, $N_T$ is the noise in the target star aperture from equation \ref{eq:ccdnoise}, and $N_E$ is the ensemble noise from equation \ref{eq:ensemblenoise}. AIJ labels the relative flux error columns as rel\_flux\_err\_T\textit{nn}, and rel\_flux\_err\_C\textit{nn} for target and comparison stars, respectively, where \textit{nn} is the aperture number.

AIJ estimates the uncertainty in each measurement based only on the factors included in equation \ref{eq:ccdnoise}. Additional sources of photometric uncertainty not accounted for by AIJ include atmospheric scintillation, variable leakage of flux from neighboring stars into the aperture as seeing changes slightly from exposure to exposure, Poisson noise in the master dark and master flat images, slight variations in CCD bias in the time-series, cosmic ray impacts on the detector, varying contributions of square pixels to a circular aperture, camera shutter open/close variations, and inaccurate determination of sky-background from exposure-to-exposure.

Atmospheric scintillation noise can dominate for short exposures and/or telescopes with small apertures. \citet{Reiger:1963} described the theoretical approach to estimating the amount of scintillation noise, \citet{Young:1967} conducted observations to test the theory and formalized the equation, and \citet{Gilliland:1993} clarified a factor in the equation. The resulting scintillation noise equation is
\begin{equation}
\sigma_{scintillation}=0.09D^{-\frac{2}{3}}\chi^{1.75}(2t_{\rm int})^{-\frac{1}{2}}e^{-\frac{h}{8000}},
\label{eq:scintnoise}
\end{equation}
\noindent where $D$ is the telescope diameter in centimeters, $\chi$ is the airmass, $t_{\rm int}$ is the exposure time in seconds, and $h$ is the altitude of the observatory in meters. Again, this noise source is \textit{not} included in the AIJ photometric uncertainty estimate.

\section{Apparent Magnitude Calculation}\label{app:apparentmagnitude}

The apparent magnitude of target aperture sources can be calculated by entering the apparent magnitude of one or more comparison aperture sources. The calculations can be done at the time multi-aperture differential photometry is being processed by enabling the \textit{Prompt to enter ref star apparent magnitude} option in the \textit{Multi-Aperture Measurements} set up panel. This option causes a \textit{Magnitude Entry} dialog box to appear when a comparison aperture is placed. The comparison aperture apparent magnitude should be entered, or the entry should be left blank if that comparison aperture is not to be used for target aperture apparent magnitude calculations. However, at least one comparison aperture apparent magnitude must be entered to enable the calculation of target aperture apparent magnitude. If WCS information is available in the FITS image header, SIMBAD information is presented by default to assist in determining the comparison star apparent magnitude. This option can be disabled in the \textit{Magnitude Entry} dialog box. Click \textit{OK} in the dialog to continue placing apertures. 

When \textit{OK} is clicked, the comparison aperture magnitude will show by default in the aperture image overlay. The first value is the magnitude entered by the user, and the second value is the net integrated counts in the aperture. The display of the two values can be controlled using the two options near the bottom of the image display \textit{View} menu in combination with the \textit{Aperture Value} display icon $\left(\,\includegraphics[scale=0.37, trim=1.5mm 2.4mm 1.5mm 0mm]{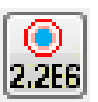}\,\right)$. The target apparent magnitudes are also calculated ``on-the-fly'' from the comparison aperture magnitudes and displayed as part of the target aperture image overlay. Each time a new comparison aperture magnitude is entered, target aperture magnitudes are recalculated from all comparison aperture magnitudes and net integrated counts.

Comparison aperture apparent magnitude values can be adjusted or added in the \textit{Multi-plot Reference Star Settings} panel after differential photometry has been completed. If magnitude entry boxes are not displayed in the panel, click the \textit{Show Magnitudes} button. After adding or changing a value, the \textit{$<$Enter$>$} key must be pressed to update the target aperture apparent magnitudes and the measurements table entries.   

SIMBAD data for a comparison aperture can be accessed by right-clicking on the corresponding comparison aperture magnitude box in the \textit{Multi-plot Reference Star Settings} panel. This feature requires that at least the first image used to derive the data in the measurements table have WCS header information, so that the RA\_\textit{Xnn} and Dec\_\textit{Xnn} columns produced by Multi-Aperture will exist with valid values in the first row of the table, where \textit{nn} is the aperture number, $X=\mathrm{T}$ for target apertures, and $X=\mathrm{C}$ for comparison apertures.

Two columns labeled Source\_AMag\_\textit{Xnn} and Source\_AMag\_Err\_\textit{Xnn} will be included in the measurements table for each target aperture and for each comparison aperture with an apparent magnitude entry, where \textit{X}  and \textit{nn} are as defined previously. The columns labeled Source\_AMag\_\textit{Xnn} contain the user entered apparent magnitudes for comparison apertures and the calculated apparent magnitudes for target apertures. The comparison aperture apparent magnitudes will be constant in the measurements table since the comparison sources are assumed to have constant brightness. The columns labeled Source\_AMag\_Err\_\textit{Xnn} contain the apparent magnitude errors. These error values do not include the uncertainty in the user entered comparison source apparent magnitudes.

If the Source\_AMag\_\textit{Xnn} columns are plotted using MP, enable the \textit{Input in Mag} option on the corresponding row of the \textit{Multi-plot Y-data} panel since these are magnitude-based data rather than flux-based data. The Source\_AMag\_\textit{Xnn} column for the target source of interest is ideal for submitting to the Minor Planet Center using the \textit{Multi-plot Main} menu option \textit{File}$\rightarrow$\textit{Create Minor Planet Center format}.

The apparent magnitude of the source in target aperture \textit{nn} is calculated as
\begin{multline}\label{eq:targetappmag}
\mathrm{Source\_AMag\_T}nn =\\
\frac{-\ln \sum_{xx} 2.512^{-\mathrm{Source\_AMag\_C}xx}}{\ln2.512}~-\\
2.5 \log{\frac{\mathrm{Source\hbox{-}Sky\_T}nn}{\sum_{xx} \mathrm{Source\hbox{-}Sky\_C}xx}}, 
\end{multline}
\noindent where \textit{nn} is the target aperture number, \textit{xx} indexes all comparison apertures for which an apparent magnitude has been entered by the user, Source\_AMag\_C\textit{xx} are the user entered comparison source apparent magnitudes, and Source-Sky\_T\textit{nn} and Source-Sky\_C\textit{xx} are the net integrated counts in apertures \textit{nn} and \textit{xx} as defined in Appendix \ref{app:measurementstable} and \S \ref{sec:photometry}. 

The uncertainty in the apparent magnitude of the source in target aperture \textit{nn} is calculated as
\begin{multline}\label{eq:targetappmagerr}
\mathrm{Source\_Amag\_Err\_T}nn = 2.5 \log\left(1 + \right.\\
\left.\sqrt{\frac{\mathrm{Source\_Error\_T}nn^{\,2}}{\mathrm{Source\hbox{-}Sky\_T}nn^{\,2}} + \frac{\sum_{xx}\mathrm{Source\_Error\_C}xx^{\,2}}{\left(\sum_{xx}\mathrm{Source\hbox{-}Sky\_C}xx\right)^2}}\right), 
\end{multline} 
\noindent where \textit{nn} is the target aperture number, \textit{xx} indexes all comparison apertures for which an apparent magnitude has been entered by the user, Source-Sky\_T\textit{nn} and Source-Sky\_C\textit{xx} are the net integrated counts and Source\_Error\_T\textit{nn} and Source\_Error\_C\textit{xx} are the  net integrated counts uncertainties for apertures \textit{nn} and \textit{xx} as defined in Appendix \ref{app:measurementstable} and \S \ref{sec:photometry}. The target aperture apparent magnitude uncertainties do not include the uncertainty in the comparison source apparent magnitudes entered by the user.

The uncertainty in the apparent magnitude of the source in comparison aperture \textit{nn} is calculated as
\begin{multline}\label{eq:compappmagerr}
\mathrm{Source\_Amag\_Err\_C}nn =\\
2.5 \log\left(1 + \frac{\mathrm{Source\_Error\_C}nn}{\mathrm{Source\hbox{-}Sky\_C}nn}\right), 
\end{multline} 
\noindent where \textit{nn} is the comparison aperture number and Source-Sky\_C\textit{nn} and Source\_Error\_C\textit{nn} are the net integrated counts and the net integrated counts uncertainty, respectively, for comparison aperture \textit{nn}. The comparison aperture apparent magnitude uncertainty is simply the flux-based photometric error converted to the magnitude scale. These values do not include the uncertainty in the comparison source apparent magnitudes entered by the user. 

\section{Measurements Tables}\label{app:measurementstable}

The results from single aperture photometry and multi-aperture differential photometry are stored in ``measurements tables". For single aperture photometry, each row in a table contains all measurements produced from a single aperture. For multi-aperture differential photometry, a row contains all measurements produced by all apertures in a single image, and a row exists for each image that has been processed. Each column contains the same measurement from all images. Columns are labeled with unique names, and those names are available for selection in the pull-down menus in the \textit{Multi-plot Main} and \textit{Multi-plot Y-data} panels. An example of part of a measurements table is shown in Figure \ref{fig:aijmeasurementstable}. As discussed in the last paragraph of Appendix \ref{sec:coordinateconverter}, new astronomical data such as AIRMASS, $\bjdtdb$, etc. can be calculated and added to the measurements table using the menu item at \textit{Multi-plot Main}$\rightarrow$\textit{Table}$\rightarrow$\textit{Add new astronomical data columns to table}. 

All data in a measurements table can be cleared by clicking the \textit{Clear Measurements Table Data} icon $\left(\,\includegraphics[scale=0.38, trim=1.5mm 2.5mm 1.5mm 0mm]{aijcleartableicon}\,\right)$ above an image display or in the \textit{Data Processor} panel. One or more rows can removed from a table by selecting them using the mouse and then selecting the menu option \textit{Edit}$\rightarrow$\textit{Clear} at the top of the measurements table. Measurements tables can be saved using the \textit{File} menu options of the measurements table and \textit{Multi-plot Main} panels. A subset of data columns from a measurements table can be saved using the menu option \textit{Multi-plot Main}$\rightarrow$\textit{Save data subset to file} as discussed in Appendix \ref{sec:saveall}. Measurements tables can be opened using the \textit{File} menu options on the \textit{Multi-plot Main} panel, the  AIJ Toolbar \textit{Read Measurements Table} icon $\left(\,\includegraphics[scale=0.37, trim=1.0mm 2.1mm 1.3mm 0mm]{aijtableopenicon}\,\right)$, or by dragging a measurements table file from an OS window and dropping it onto any of the Multi-Plot control panels. Any tab, comma, or space delimited data table can be opened as a measurements table. However, we strongly suggest that the first row of the table contain text-based labels for each column so that the Multi-Plot data column labels will be representative of the data in each column of the table. 

The measurements and associated column labels produced when preforming multi-aperture differential photometry are described in the list below. A suffix ``\_\textit{Xnn}'' denotes the aperture identification, where \textit{X} is either T or C for target or comparison aperture, respectively, and \textit{nn} is the aperture number. Each measurement item can be selected for inclusion in the measurements table in the \textit{More Aperture Photometry Settings} panel (see Appendix \ref{sec:photset}). 

\begin{figure*}
\begin{center}
\resizebox{\textwidth}{!}{
\includegraphics{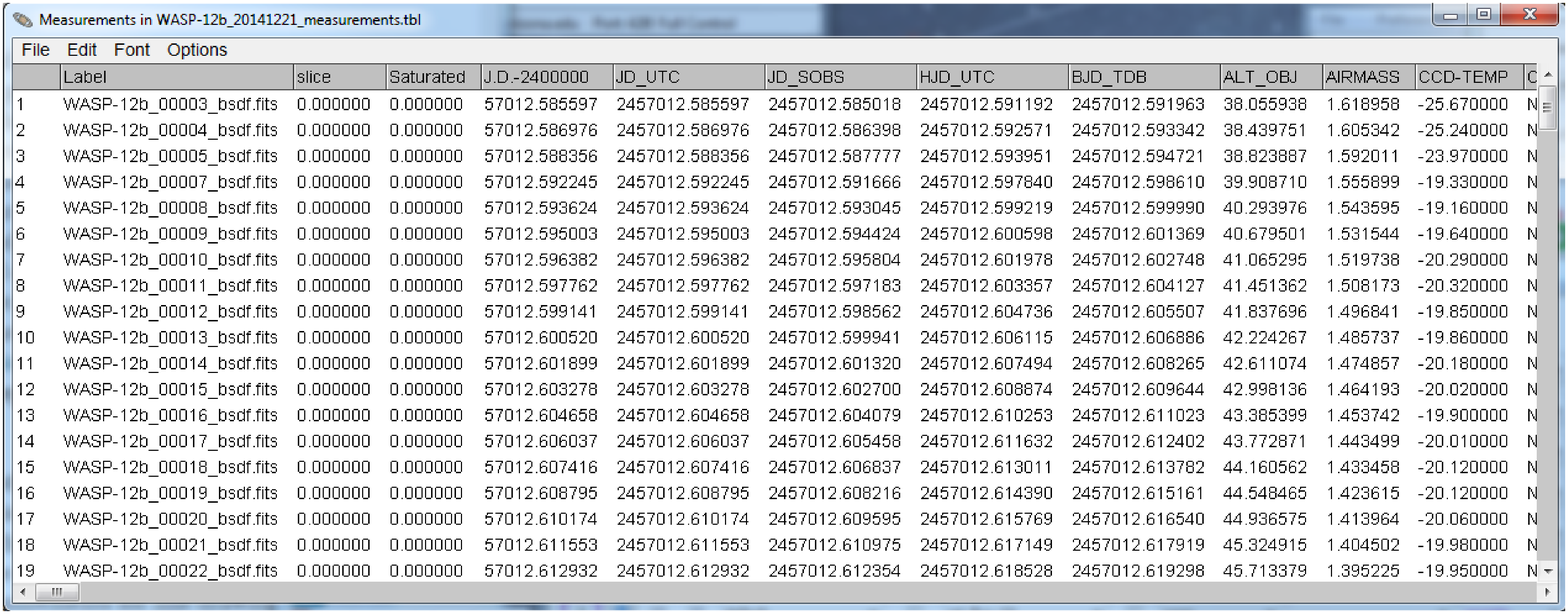} }
\caption{Part of a Measurements Table. Multi-aperture differential photometry produces a measurements table containing all photometric data from the time series. Each row contains all measurements from an individual image. Each column contains the same measurement from all images. Columns are labeled with unique names, and those names are available for selection in the pull-down lists in the \textit{Multi-plot Main} and \textit{Multi-plot Y-data} panels.} 
\label{fig:aijmeasurementstable}
\end{center}
\end{figure*}

\begin{itemize}
 
\item \textbf{J.D.-2400000}: modified Julian Date of the observation (UTC based). Availability requires UTC time and date information in the FITS header. If keyword EXPTIME is also available in the header, time of mid-exposure is indicated, otherwise start of exposure is indicated.  
\item \textbf{JD\_UTC}: If this keyword value is set and successfully extracted from the FITS header, the header value is indicated. Otherwise, the full Julian Date of the observation (UTC based) is indicated based on the value in J.D.-2400000. 
\item \textbf{Source\_Radius}: radius of the aperture used to calculate net integrated counts. In fixed aperture mode, this value is the aperture radius set by the user. In variable aperture mode, this value represents the actual aperture radius calculated as the product of the average FWHM in the image and the multiplicative factor set in the \textit{Multi-Aperture Measurements} set up panel.
\item \textbf{FWHM\_mult}: in variable aperture mode, this value is the FWHM multiplier set in the \textit{Multi-Aperture Measurements} set up panel. In fixed aperture mode, this column is not included in the table.  
\item \textbf{Source\_Rad(base)}: in variable aperture mode, this value represents the fixed aperture radius set by the user and should be set to a number greater than 1.5 times the maximum FWHM expected to ensure proper measurement of FWHM. In fixed aperture mode, this column is not included in the table.  
\item \textbf{Sky\_Rad(min)}: radius of the inner edge of the annulus used to calculate the sky background  
\item \textbf{Sky\_Rad(max)}: radius of the outer edge of the annulus used to calculate the sky background  
\item \textbf{X(IJ)\_\textit{Xnn}}: the x-location in ImageJ pixel coordinates of the center of aperture \textit{nn} 
\item \textbf{Y(IJ)\_\textit{Xnn}}: the y-location in ImageJ pixel coordinates of the center of aperture \textit{nn}   
\item \textbf{X(FITS)\_\textit{Xnn}}: the x-location in FITS coordinates of the center of  aperture \textit{nn}  
\item \textbf{Y(FITS)\_\textit{Xnn}}: the y-location in FITS coordinates of the center of  aperture \textit{nn}  
\item \textbf{rel\_flux\_T\textit{nn}}: the ratio of the net integrated counts in target aperture \textit{nn} to the total net integrated counts of all comparison stars. Mathematically the value is calculated as Source-Sky\_T\textit{nn} / $\sum$(Source-Sky\_C\textit{xx} ), where \textit{xx} indexes all comparison star apertures.  
\item \textbf{rel\_flux\_C\textit{nn}}: the ratio of the net integrated counts in comparison aperture \textit{nn} to the total net integrated counts of all \textit{other} comparison stars. Mathematically the value is calculated as Source-Sky\_C\textit{nn} / $\sum$(Source-Sky\_C\textit{xx}), where \textit{xx} indexes all comparison star apertures and $xx\ne nn$.
\item \textbf{rel\_flux\_err\_\textit{Xnn}}: the error in the relative flux for object \textit{nn}. The error is calculated by propagating all Source\_Error\_\textit{Xnn} values through the equations defined for rel\_flux\_\textit{Xnn} as described in Appendix \ref{app:photerr}.  
\item \textbf{rel\_flux\_SNR\_\textit{Xnn}}: signal-to-noise ratio for rel\_flux\_\textit{Xnn}. This value is simply rel\_flux\_\textit{Xnn} / rel\_flux\_err\_\textit{Xnn}.  
\item \textbf{Source-Sky\_\textit{Xnn}}: net integrated counts within aperture \textit{nn}. Net integrated counts is defined as the sum of all pixel values less sky background for all pixels having a center that falls within the aperture radius.
\item \textbf{Source\_Error\_\textit{Xnn}}: error in the net integrated counts for aperture \textit{nn}. See Appendix \ref{app:photerr} for a description of how the photometric error is calculated.  
\item \textbf{Source\_SNR\_\textit{Xnn}}: signal-to-noise ratio aperture \textit{nn}. This value is simply Source-Sky\_\textit{Xnn} / Source\_Error\_\textit{Xnn}.  
\item \textbf{tot\_C\_cnts}: the sum of the net integrated counts in all comparison apertures. tot\_C\_cnts = $\sum$(Source-Sky\_C\textit{xx}), where \textit{xx} indexes all comparison apertures. 
\item \textbf{tot\_C\_err}: the error in tot\_C\_cnts. The error is calculated by combining all Source\_Error\_C\textit{xx} values in quadrature, where \textit{xx} indexes all comparison stars.  
\item \textbf{Peak\_\textit{Xnn}}: the highest pixel value in aperture \textit{nn} (\textit{not} background subtracted)  
\item \textbf{Mean\_\textit{Xnn}}: the mean pixel value in aperture \textit{nn} (\textit{not} background subtracted)
\item \textbf{Sky/Pixel\_\textit{Xnn}}: the sky background estimate for aperture \textit{nn}. Depending on the options selected, the value may be the mean value of all pixels, the mean value of all pixels after iterative $2\sigma$ cleaning, or a plane fitted to (possibly $2\sigma$ iteratively cleaned) pixels in the sky-background annulus of aperture \textit{nn}.
\item \textbf{X-Width\_\textit{Xnn}}: an estimate of the x-direction FWHM of the PSF of the object in aperture \textit{nn}. The aperture radius should extend beyond 1.5 $\times$ FWHM in the x-direction to enable the best estimate.
\item \textbf{Y-Width\_\textit{Xnn}}: an estimate of the y-direction FWHM of the PSF of the object in aperture \textit{nn}. The aperture radius should extend beyond 1.5 $\times$ FWHM in the y-direction to enable the best estimate.
\item \textbf{Width\_\textit{Xnn}}: the mean of the x- and y-direction FWHM values for aperture \textit{nn}. The aperture radius should extend beyond 1.5 $\times$ FWHM in the both the x- and y-directions to enable the best estimate.
\item \textbf{Saturated}: zero indicates that the peak value in all apertures is less than the \textit{Saturation warning} value set in the \textit{Aperture Photometry Settings} panel (see Appendix \ref{sec:photset}). A number other than zero indicates the highest peak value that exceeded the set limit. 
\item \textbf{Source\_AMag\_C\textit{nn}}: comparison aperture \textit{nn} apparent magnitude entered by the user. This column's values are the same in all rows since the comparison source is assumed to be constant. This column does not exist if apparent magnitude has not been entered for aperture \textit{nn}.
\item \textbf{Source\_AMag\_T\textit{nn}}: target aperture \textit{nn} apparent magnitude calculated from all comparison apertures having an apparent magnitude entry. This column does not exist if apparent magnitude has not been entered for any comparison apertures. Equation \ref{eq:targetappmag} defines how this value is calculated.
\item \textbf{Source\_AMag\_Err\_T\textit{nn}}: target aperture \textit{nn} apparent magnitude uncertainty. This column does not exist if apparent magnitude has not been entered for any comparison apertures. Equation \ref{eq:targetappmagerr} defines how this value is calculated.
\item \textbf{Source\_AMag\_Err\_C\textit{nn}}: comparison aperture \textit{nn} apparent magnitude uncertainty. This column does not exist if apparent magnitude has not been entered for aperture \textit{nn}. Equation \ref{eq:compappmagerr} defines how this value is calculated.
\end{itemize}

The numeric values of FITS header keywords can also be extracted from an image (or series of images) and included in a measurements table when performing single aperture or multi-aperture differential photometry. The associated column in the table will be labeled using the keyword name. These keyword names must be specified in the \textit{Keywords (comma separated)} list in the \textit{Aperture Photometry Settings} panel (see Appendix \ref{sec:photset}). As discussed in Appendix \ref{sec:fitsheaderupdates}, DP provides the option to calculate new astronomical data and add them to the FITS header of a calibrated image. A keyword name specified in the \textit{Keywords (comma separated)} list in the \textit{Aperture Photometry Settings} panel must match the keyword name specified in the \textit{FITS Header Output Settings} sub-panel of the \textit{General FITS Header Settings} panel to properly transfer the DP calculated astronomical data into the measurements table. In addition, the \textit{FITS Header Update} option must be enabled and the target and observatory coordinates must be provided through one of the various options in the DP \textit{FITS Header Updates} sub-panel. UTC date, time, and exposure time information must also exist in the FITS header of the raw image. Some keywords containing astronomical data calculated by DP that are often extracted from the calibrated image headers and added to the measurements table are listed below.

\begin{itemize}

\item \textbf{HJD\_UTC}: UTC-based heliocentric Julian Date of this measurement at the mid-point of the exposure (see \citealt{Eastman:2010} for a complete definition of $\mathrm{HJD}_{\mathrm{UTC}}$)
\item \textbf{BJD\_TDB}: TDB-based barycentric Julian Date of this measurement at the mid-point of the exposure (see \citealt{Eastman:2010} for a complete definition of $\mathrm{BJD}_{\mathrm{TDB}}$)
\item \textbf{ALT\_OBJ}: altitude of the target coordinates above the horizon at mid-exposure
\item \textbf{AIRMASS}: airmass of the target coordinates at mid-exposure 

\end{itemize}

\bibliographystyle{apj.bst}

\bibliography{AstroImageJ.bib}

\end{document}